\documentclass[aps,prd,superscriptaddress,showpacs,nofootinbib]{revtex4}

\usepackage{amssymb,amsmath,amsfonts,amsthm,graphicx}

\usepackage{color}
\usepackage{graphicx}

\def\be{\begin{equation}}
\def\ee{\end{equation}}
\newcommand{\bea}{\begin{eqnarray}}
\newcommand{\eea}{\end{eqnarray}}
\newcommand{\lum}{\mathcal{L}}

\begin{document}

\title{Cosmological Radio Emission induced by WIMP Dark Matter}
\thanks{Preprint number: DFTT 33/2011}

\author{N. Fornengo}
\affiliation{Dipartimento di Fisica Teorica, Universit\`a di Torino, I-10125 Torino, Italy}
\affiliation{Istituto Nazionale di Fisica Nucleare, Sezione di Torino I-10125 Torino, Italy}
\author{R. Lineros}
\affiliation{IFIC, CSIC--Universidad de Valencia, Ed. Institutos, Apdo. Correos 22085, EÐ46071 Valencia, Spain and MultiDark fellow}
\author{M. Regis}
\affiliation{Dipartimento di Fisica Teorica, Universit\`a di Torino, I-10125 Torino, Italy}
\affiliation{Istituto Nazionale di Fisica Nucleare, Sezione di Torino I-10125 Torino, Italy}
\author{M. Taoso}
\affiliation{IFIC, CSIC--Universidad de Valencia, Ed. Institutos, Apdo. Correos 22085, EÐ46071 Valencia, Spain and MultiDark fellow}
\affiliation{Department of Physics and Astronomy, University of British Columbia, Vancouver, BC V6T 1Z1, Canada}

\begin{abstract}
We present a detailed analysis of the radio synchrotron emission induced by WIMP dark matter annihilations and decays in extragalactic halos.
We compute intensity, angular correlation, and source counts and discuss the impact 
on the expected signals of dark matter clustering, as well as of other astrophysical uncertainties as magnetic fields and spatial diffusion. 
Bounds on dark matter microscopic properties are then derived, and, depending on the
specific set of assumptions, they are competitive with constraints from other indirect dark matter searches.
At GHz frequencies, dark matter sources can become a significant fraction of the total number of sources with brightness below the microJansky level.  We show that,
at this level of fluxes (which are within the reach of the next--generation radio surveys), properties of the faint edge of differential source counts, as well as angular correlation data, can become an important probe for WIMPs.
\end{abstract}

\maketitle

\section{Introduction}
The nature of non--baryonic Dark Matter (DM) is one of the most puzzling mysteries of modern Cosmology (see e.g. \cite{Bertone:2004pz,Bergstrom:2000pn} for reviews).
A non--gravitational detection of dark matter would be a fundamental step toward the comprehension of its properties.
This could be achieved in several ways, the most promising and popular methods are
direct searches in underground experiments, DM searches at colliders, and indirect detection methods.
The latter techniques are based on the detection of the products of
DM annihilations or decays, notably neutrinos, antimatter and photons.
Indirect searches are particularly promising if dark matter is in the form of Weakly Interacting Massive Particles (WIMPs).
In the early Universe, these particles decouple non--relativistically from the thermal plasma and inherit the correct relic
abundance for a thermally averaged annihilation cross section 
close to $\langle\sigma v\rangle = 3\times 10^{-26} \mbox{ cm}^3\mbox{s}^{-1}.$
Annihilations of WIMPs inside  galaxies are able to inject large amounts of relativistic electrons and positrons, which in turns
interact with the galactic magnetic fields producing synchrotron radiation.
This is a generic prediction of WIMP models, except for the peculiar case of WIMP candidates annihilating only into neutrinos.
For galactic magnetic fields of the order of $\mu$G and electrons energies below about 10 GeV, the synchrotron
emission falls at frequencies around and below the GHz , i.e. in the radio band.
Therefore, radio observations are appropriate tools for indirect WIMP searches and they have been extensively
discussed in the context of WIMP annihilations inside our 
galaxy \cite{Borriello:2008gy,Boehm:2010qt,Bertone:2002je,Boehm:2010kg,Crocker:2010gy,Regis:2008ij,Bergstrom:2008ag,Bertone:2008xr,Cumberbatch:2009ji,Dobler:2007wv,Linden:2010eu,Hooper:2007kb,Fornengo:2011iq}.
In particular, low frequencies ($\lesssim $ 1 GHz) data severely constraint
WIMP masses $\lesssim$ 10 GeV for leptonic annihilation modes \cite{Fornengo:2011iq}.

Here we focus instead on the radio emission produced by DM annihilations or decays in extragalactic halos.
A previous analysis along this direction can be found in Ref.~\cite{Zhang:2008rs}.
Recently, we have shown that for realistic assumptions on the DM clustering and magnetic fields
and for a ``thermal'' annihilation cross--section (i.e. $(\sigma v)=3\times 10^{-26} \mbox{ cm}^3\mbox{s}^{-1}$), the DM
signal could account for a significant fraction of the extragalactic--radio emission inferred by the ARCADE 2 Collaboration \cite{Fornengo:2011cn}.
This constitutes a further motivation to seriously consider DM searches with extragalactic radio observations.

In this work, we extend the studies in \cite{Zhang:2008rs,Fornengo:2011cn} by extensively analyzing the impact of 
astrophysical and particle physics uncertainties on the DM extragalactic radio emission which is then compared to observational data.
We focus on three observables: intensity, differential number counts of sources and angular correlations. 
By comparing predictions with current data, constraints on the DM annihilation/decay rate and mass are derived.
We also compare the DM signals with (data--driven) expectations for astrophysical sources, such as radio loud active galactic nuclei
and star forming galaxies, and show that source counts and angular correlation data could be particularly relevant
to distinguish the DM contribution from the astrophysical ones.
 
It is interesting to note that cosmological radio emission induced by WIMP DM decreases very rapidly with redshift 
(so the main contributions come from structures at $z\ll1$), and that DM halos can substantially contribute 
to the counts of sources at brightness below the $\mu\mbox{Jy}$ level.
Therefore, a specific property of the extragalactic WIMP--induced source population is to peak at low redshift and low brightness.
This hypothesis can be tested by telescopes with flux sensitivity at $\mu\mbox{Jy}$ level, 
which will be reached by upcoming radio experiments, in particular by the
Square Kilometer Array (SKA).  

The paper is organized as follows.
In Section \ref{sec:obsdata} we review the present status of radio observations, focusing
on those particularly relevant for DM searches. We discuss data on intensity, angular correlations
and counts of sources. The formalism needed to compute the DM and astrophysical signals is presented in Section \ref{sec:thmod}.
Section \ref{sec:bench} shows our results for both DM and astrophysical models. Our estimates are
first derived focusing on few specific DM benchmark cases while a detailed analysis about
the impact of astrophysical and particle physics uncertainties is presented in
Section \ref{sec:disc}. The bounds on the microscopic properties of DM are computed
in Section \ref{sec:constraints}, and we derive our conclusions in Section \ref{sec:conclusions}.

\section{Extra--galactic radio signals}

\subsection{Observational data}
\label{sec:obsdata}

\subsubsection{Total intensity of cosmic background}
Although the history of radio observations is quite long and rich, only few maps have sufficient angular resolution and sky coverage to allow estimates of the isotropic extragalactic background. In particular, they need to be suitable to reliably separate the large--scale (astrophysical) Galactic component, which is a delicate procedure.

Recently, the ARCADE 2 (Absolute Radiometer for Cosmology, Astrophysics and Diffuse Emission) project~\cite{Singal:2009xq} reported the measurement of the extra--galactic sky temperature  at frequencies ranging from 3 to 90 GHz~\cite{Fixsen:2009xn,Seiffert:2009xs}. At high frequency, the radiation is largely dominated by the CMB, so the estimate of other extra--galactic components becomes unfeasible. At frequencies down to GHz the CMB is still dominant but its contribution can be easily subtracted since it follows a black--body (BB) spectrum and we know its BB temperature with per--mil precision. At lower frequencies, extra--galactic sources dominate.

In the ARCADE data, the diffuse Galactic foreground is separated using two independent estimators (namely, a co--secant dependence on Galactic latitude and the correlation between radio and atomic line emissions), which agree well between each other \cite{Kogut:2009xv}. In Ref.~\cite{Fixsen:2009xn} similar analyses have been performed on high--latitude patches in past surveys at 22, 45, 408, and 1420 MHz. 
We consider the estimates derived with this procedure as a conservative constraint for the extra--galactic signal since they might include a contribution from a (isotropic) Galactic emission uncorrelated with th gas (e.g., from the DM halo).
Data--points shown throughout the paper are taken from Table~4 in Ref. \cite{Fixsen:2009xn} by converting the thermodynamic temperature $T_0$ to the brightness temperature $T_B$ by means of the usual procedure: $T_B=T_0\cdot x/(e^x-1)$ with $x=h\nu/(kT_0)$.

The expected signal of astrophysical extragalactic sources derived from models which fit source counts (see next Section) lies well below the data: this
points towards the existence of a previously disregarded component. The extragalactic (or Galactic) DM emission can account for this excess and this possibility has been explored in
Ref. \cite{Fornengo:2011cn}.

We will also consider estimates of the cosmic infrared background derived in Ref. \cite{Gispert:2000np} (with CMB subtracted) and based on FIRAS and DIRBE datasets.

\subsubsection{Angular correlation}
The angular distribution of sources in the sky is a powerful probe of large--scale clustering.
Probing the latter (which includes disentangling contributions from different redshift slices) at radio frequencies is rather hard, but,
nevertheless, wide--area radio surveys are a useful tool, since they can test clustering on the largest scales. 
Recently, observational campaigns (e.g., FIRST~\cite{Becker:1995ei} and NVSS~\cite{Condon:1998iy}), have been exploited for this purpose.

Angular distributions are often reported in terms of the two--point angular correlation function $\omega(\theta)$ or the angular power spectrum $C_\ell$.
Theoretically, such two estimators are fully equivalent and simply related by a Legendre transformation:
\be
C_\ell=\frac{2\pi\,N^2}{\Delta \Omega^2}\int^1_{-1}d(\cos{\theta})\,\omega(\theta)\,P_\ell(\cos{\theta})\;\;\; {\rm or}\;\;\;
\omega(\theta)=\frac{\Delta \Omega^2}{4\pi\,N^2}\sum_{\ell=1}^{\infty}\,(2\,\ell+1)\, C_\ell\,P_\ell(\cos{\theta})\;,
\label{eq:wcl}
\ee
where $N$ is the number of sources, $\Delta \Omega$ is the survey area, and $P_\ell$ are the Legendre polynomials.
On the other hand, $C_\ell$ and $\omega(\theta)$ quantify different properties of the angular distribution and so they are not observationally equivalent (i.e., to perform the transformation we would need data at all angles, which is of course unrealistic).
Naively, $C_\ell$ is a better probe at large angles (provided a sufficiently large sky coverage), while $\omega$, which suffers from Poisson noise on large scales, is a superior tool for small angular separations. 

For what concerns the angular two--point correlation function of extragalactic sources, the state of the art bears on flux thresholds of few mJy and angular separation of (fraction of) arcmin.
Since the DM contribution is faint, we aim at analyzing correlation functions with flux threshold as low as possible.
We consider data from the FIRST survey~\cite{Cress:1996vz} at 1.4 GHz analyzed in Ref.~\cite{Overzier:2003kg} (their Fig.~3), which concerns clustering of sources with $S>3$ mJy. For our purposes, although other measurements could be considered (including surveys of VLSS at 74 MHz~\cite{deOliveiraCosta:2009wu}, WENSS at 325 MHz~\cite{Blake:2003dv}, SUMSS at 843 MHz~\cite{Blake:2003dv}, and NVSS at 1.4 GHz~\cite{Blake:2001bg}), they do not add further insights since they have similar (or worse) flux thresholds.
We anticipate already here that the number of DM sources with mJy brightness at GHz frequencies is very low for typical WIMP models, thus the clustering tested by those surveys is not really relevant to DM.
To partially overcome such issue one can consider measurements of angular correlation functions in small but deep fields (e.g., at 1.4 GHz in~\cite{Wilman:2002zh} with $S>0.2$ mJy). However, in this case the statistics is much poorer and this makes any interpretation significantly harder.
As we will discuss later, flux thresholds at $\mu$Jy level would be in order to possibly probe realistic WIMP scenarios, and this can be actually achieved only by future surveys with SKA.
The angular power spectrum $C_\ell$ of the radio source distribution can be inferred (under some assumptions and using Eq.~\ref{eq:wcl}) from the angular correlation function. This has been performed in Ref.~\cite{Blake:2004dg} for multipoles below $\ell\sim100$ analyzing  NVSS data at 1.4 GHz. However, large statistical errors and, again, high flux thresholds limit their usefulness for constraining WIMP models.

Radio $C_\ell$ of anisotropies are mainly extracted from sky--maps of single--dish radio telescopes at low--frequency and from CMB--oriented experiments at high--frequency. 
The firsts have typically low angular resolution, and provide estimates of $C_\ell$ up to $\ell$ of few hundreds, while the Galactic foreground dominates up to $\ell\sim100$. We consider angular power--spectra at different frequencies derived in Ref.~\cite{Regis:2011ji} with estimates of source contribution as in Ref.~\cite{Giardino,LaPorta:2008ag}. We derive a band which brackets uncertainties related to the disentanglement between extragalactic and galactic foreground anisotropies.

At CMB frequencies, the CMB fluctuations dominates over extra--galactic sources except for $\nu\gtrsim100$ GHz and $\ell\gtrsim3000$ (where unfortunately IR sources constitute the main contribution). We consider recent data at 150 GHz from SPT~\cite{Reichardt:2011yv} (similarly, the analysis of ACT data~\cite{Dunkley:2010ge} leads to analogous conclusions).

Therefore, at present, measurements of radio angular correlation do not provide very powerful constraints on non--gravitational DM signals. On the other hand, our analysis is mainly oriented towards understanding anisotropies signatures of WIMPs and how they could be tested by future surveys.

\subsubsection{Source number counts}
Source number counts are an important tool to understand properties of different radio populations.
Although counting sources of a given brightness may sound rather trivial, it is well known that 
 obtaining a population count from a sky survey is not an easy task, and different telescopes 
(endowed with, e.g., different angular resolution and sensitivity) introduce different biases that need to be properly accounted for.
A wide consensus has been reached nowadays on the interpretation that radio counts are considered to be dominated by radio--loud AGNs down to mJy--levels, 
while star--forming galaxies (and possibly radio-quiet AGNs) take over at fainter flux--density. For a recent review, see, e.g., Ref.~\cite{DeZotti:2009an}.

The number of sources per unit area $N$ is often reported in terms of the differential counts $dN/dS$ (where $S$ is the flux density) 
and normalized to the Euclidean case, i.e., multiplied by $S^{5/2}$ (a uniform source distribution in a static Euclidean universe leads to $N \propto S^{-3/2}$, since the number of sources is proportional to the volume, while the flux density scales as the inverse distance squared).

A collection of differential source counts ranging from 150 MHz to 20 GHz is provided in Ref.~\cite{DeZotti:2009an} (see references therein for each dataset), and will be exploited in this work.

\subsection{Theoretical models}
\label{sec:thmod}
Considering an extragalactic source population with number density $n$ and with luminosity $\lum$ depending on a certain quantity $Q$, the total isotropic intensity per solid angle at a given frequency $\nu$ is given by~\cite{Ando:2005xg,Zhang:2008rs}:
\bea
\nu\,I_\nu&=&\nu\,\int dz \frac{d^2V}{dzd\Omega}\int dQ \frac{dn}{dQ}(Q,z)\frac{(1+z)\,\lum(E=E_\nu(1+z),z,Q)\,e^{-\tau(z)}}{4\pi\,d_L(z)^2} \nonumber \\
&=&\frac{c\,\nu}{4\pi}\int dz\frac{e^{-\tau(z)}}{(1+z)\,H(z)}\int dQ \frac{dn}{dQ}(Q,z)\,\lum(E=E_\nu(1+z),z,Q)\;, \label{eq:intgen}
\eea
where $V$ is the comoving volume, $\Omega$ is the solid angle, $z$ is the redshift, $E$ is the emission energy, $d_L$ is the luminosity distance, $H$ is the Hubble rate, and $\tau$ is the optical depth.
$Q$ will be $M$ (halo mass) for DM and $\lum_{\nu_0}$ (luminosity at a reference frequency $\nu_0$) for astrophysical sources.

The differential number counts of sources per solid angle can be written as:
\be
\frac{dN}{dS_\nu}= 4\pi\,c \int dz \frac{e^{-\tau(z)}\,d_L(z)^4}{(1+z)^3\,H(z)}\frac{dn}{dQ}(Q,z)\,\frac{dQ}{d\lum_{\nu'}}\;,
\label{eq:defN}
\ee
where $S_\nu=(1+z)\,\lum_{\nu'}/(4\pi\,d_L(z)^2)$ and $\nu'=(1+z)\,\nu$.

Eqs.~(\ref{eq:intgen}) and (\ref{eq:defN}) can be simply linked through:
\be
I_\nu= \int_0^{S_0} dS \frac{dN}{dS_\nu}S\;,
\ee
where $S_0$ is a minimum (telescope--dependent) apparent source--brightness above which sources can be detected by the telescope and so subtracted from the total isotropic flux (we will mostly refer to $I_\nu$ as the contribution from unresolved extragalactic sources).

The angular power spectrum of cosmic background anisotropies is~\cite{Ando:2005xg,Zhang:2008rs}:
\bea
C_\ell&=&C_\ell^{1h}+C_\ell^{2h}\label{eq:cl}\\
C_\ell^{1h}&=&\frac{c}{(4\pi)^2}\int dz\frac{e^{-2\tau(z)}}{d_L(z)^2\,H(z)}\int dQ \frac{dn}{dQ}(Q,z)\,[\lum(E=E_\nu(1+z),z,Q)\,|u(k|Q)|]^2 \label{eq:cl1h}\\
C_\ell^{2h}&=&\frac{c}{(4\pi)^2}\int dz\frac{e^{-2\tau(z)}}{d_L(z)^2\,H(z)}P^{lin}(k,z)\left[\int dQ \frac{dn}{dQ}(Q,z)\,\lum(E=E_\nu(1+z),z,Q)\,b(Q,z)\,|u(k|Q)|\right]^2\;, \label{eq:cl2h}
\eea
where $u$ is the Fourier transform of the spatial emission density profile of the source (with the latter normalized such that its volume integral is equal to unity), and $b$ is the linear bias factor that relates the source clustering to the matter clustering.
Label ``$1h$'' denotes correlations between particles within the same halo, while ``$2h$'' describes correlations between particles in two distinct halos (and since we are interested in the two--point correlation, there are no further terms).

The Limber equation provides the relation between number counts and angular power spectrum or correlation function of a given class of objects~\cite{Limber,Peebles}.
In particular, we will make use of (see, e.g., Ref.~\cite{Eisenstein:1999jg}):
\be
\omega(\theta)=\int_0^\infty dk\,k\,\int_0^\infty dz\,P(k,z)\,\frac{J_0(k\,\theta\,r)}{2\pi}\,\left(\frac{1}{N}\frac{dN}{dz}\right)^2 \frac{dz}{dr}\;,
\label{eq:wdef}
\ee
where $P(k,z)$ is the 3D power--spectrum, $J_0$ is the zero--order Bessel function, $dz/dr=H(z)/c$ in a flat universe, and $dN/dz$ can be derived from Eq.~(\ref{eq:defN}).
Note also that, in the idealized case of uniformly distributed point--sources, $C_\ell$ does not vary with $\ell$ (i.e., it is a Poisson noise term) and is simply given by $C_\ell=\int_0^{S_0} dS\,dN/dS\,S^2$.

Absorption along the line of sight (described through the optical depth $\tau$) is negligible at radio frequencies above $\sim10$ MHz and will be disregarded.
Possible absorption effects within the source (e.g., synchrotron self--absorption) are instead included in $\lum$; however, for the (non--compact) DM sources considered here, they are negligible as well.
 
\subsubsection{Dark matter source}
We consider the DM to be in form of particles (with mass $M_{\chi}$) which can annihilate (with annihilation rate ($\sigma_a v$)) or decay (with decay rate $\tau_d^{-1}$) into (among other products) electrons and positrons. The latter in turn emit photons through synchrotron radiation.
Up to the injection of $e^-/e^+$ and for a given DM model, the process is completely set by the amount of DM present in the source, which implies $Q=M$, where $M$ is the DM halo mass.
On the other hand, some of the quantities involved in the diffusion and energy loss of $e^-/e^+$ after production (e.g., magnetic field) could be not strictly correlated with the mass of the source, but rather to properties of host galaxies, merger history, etc. 
As a first approximation, however, their possible variations from source to source can be reasonably associated only to halo mass and redshift, and so, for simplicity, we will assume $Q=M$.
The luminosity $\lum$ of DM sources is given by
\bea
\lum_a^{hh}(E,z,M)&=&E\,\frac{(\sigma_a v)}{2\,M_{\chi}^2} \int_0^{R_v} d^3r \frac{d\tilde N_i}{dE}\,[(1-f)\,\rho(M,r,z)]^2\hspace{1cm}{\rm for\,the\,host\,halo}\label{eq:lumDMa1}\\
\lum_a^{sh}(E,z,M)&=&E\,\frac{(\sigma_a v)}{2\,M_{\chi}^2} \int_{M_{cut}^s}^M dM_s\frac{dn_s}{dM_s}(M_s,f,M)\int_0^{R_v} d^3r_s \frac{d\tilde N_i}{dE}\,\rho_s^2(M_s,r_s,z)\hspace{1cm}{\rm for\,subhalos}
\label{eq:lumDMa}
\eea
in the annihilating DM case, and by
\be
\lum_d(E,z,M)=\frac{E}{\tau_d\,M_{\chi}} \int_0^{R_v} d^3r \frac{d\tilde N_i}{dE}\,\rho(M,r,z)
\label{eq:lumDMd}
\ee
in the decaying DM case.
We introduced $\rho$ which is the total DM density profile (cut at virial radius $R_v$), and $\rho_s$ which is the density profile of a single clump. 
Note that in Eq.~\ref{eq:lumDMa} there is no dependence on the spatial distribution of clumps  within the host halo. This is not relevant when calculating the total isotropic 
intensity, while the information is encoded in $u$ for the angular power spectrum.
The spatial profile of the main halo is given by $\rho_h=(1-f)\,\rho$. Thus $f$ provides the fraction of total halo mass given by substructures with mass function $dn_s/dM_s$ (normalized such that $\int dM_s\,dn_s/dM_s\,M_s=f\,M$). The minimum subhalo mass is $M_{cut}^s$, and $d\tilde N_i/dE=d\tilde N_i/dE(E,r,z)$.
We aim at providing equations suitable for multi--wavelength analyses and the subscript $i$ denotes different emission mechanisms; more precisely
\bea
\frac{d\tilde N_\gamma}{dE}&=&\frac{dN_\gamma}{dE}\;,\;\;{\rm for \;prompt\;emission}\\
\frac{d\tilde N_{syn,IC}}{dE}&=&2\,\int^{M_{\chi}}_{m_e}dE'\, \frac{P_{syn,IC}}{E}\cdot \tilde n_e\;,\; {\rm for\;radiative\; emission}\;.
\eea
where $\tilde n_e(r,E)=n_e/A$, with $A_a=(\sigma v)/2\cdot(\rho/M_{\chi})^2$ and $A_d=\tau_d^{-1}\,\rho/M_{\chi}$, and $n_e$ being the electron/positron equilibrium number density obtained solving a transport equation for $e^-/e^+$ injected by DM with an energy spectrum set by $dN_e/dE$ (see, e.g., review in Ref.~\cite{Profumo:2010ya}). The function $dN_j/dE$ with $j=\gamma,e$ describes spectra of annihilation/decay into gamma--rays and electrons/positrons, respectively.
The radiative emission power in the synchrotron case is given by:
\begin{equation}
 P_{syn} (r,E,\nu)= \frac{\sqrt{3}\,e^3}{m_e c^2} \,B(r) F(\nu/\nu_c)
\end{equation}
where $m_e$ is the electron mass, the critical synchrotron frequency is defined as $\nu_c \equiv  3/(4\,\pi) \cdot {c\,e}/{(m_e c^2)^3} B(r) E^2$,  and $F(t) \equiv t \int_t^\infty dz K_{5/3}(z)$ is the function setting the spectral behavior of synchrotron radiation. For the
inverse--Compton (IC) process we have:
\begin{equation}
P_{IC}(r,E,\nu) = c\,h\nu \int d\epsilon\, n_\gamma(\epsilon,r)\,\sigma(\epsilon,\nu,E)
\label{eq:PIC}
\end{equation}
where $\epsilon$ is the energy of the target photons, $n_\gamma$ is their energy
spectrum, and  $\sigma$ is the Klein--Nishina cross section.

The quantity $u$ in Eqs.~\ref{eq:cl1h} and \ref{eq:cl2h} is the Fourier transform of the spatial emission density profile $U$ of the source normalized to unity (i.e., $\int d^3r\,U=1$). In the annihilating case, for the main halo, $U$ is proportional to $\rho_h^2$, and turns out to be simply given by $A_a\cdot E/\nu\cdot d\tilde N_i/dE$ up to the normalization factor. In the substructure case, we consider a scenario in which $U$ follows from $\rho_h$ of the parent halo (biased), and an antibiased case modeled from results in~\cite{Madau:2008fr} (which are assumed to hold for any M ) where the subhalos are more preferentially located in the outer part of the parent halo. With this treatment, clumps are effectively considered as point sources with spatial probability set by $U$, thus disregarding the actual clumpiness of subhalo distribution which however would affect only angular correlations at extremely small angles.\footnote{Our approach is fully equivalent to the description of Ref.~\cite{Ando:2006cr}, by rewriting their Halo Occupation Distribution as $\langle N|M \rangle=\int dM_s dn/dM_s \int d^3r_s \rho_s^2$ which implies $\lum_a^{sh}=E_z (\sigma_a v)/(2M_{\chi}^2)\,d\tilde N_i/dE \,\langle N|M\rangle$.}

For decaying models, the spatial emission density profile of the source is given by $A_d\cdot E/\nu\cdot d\tilde N_i/dE$ with $\rho=\rho_h+\rho_s$ in the definition of $A_d$. 

Neglecting spatial dependencies in $d\tilde N_i/dE$, this term goes out of the volume integral in Eqs.~\ref{eq:lumDMa1}, \ref{eq:lumDMa}, and \ref{eq:lumDMd}.
This implies also that $\tilde n_e$ is spatially constant which can be satisfied only if in the transport equation for $n_e$ there is no spatially dependent term (except for the source term itself which in this case encodes all the spatial information). In particular, under this approximation, electrons and positrons radiate at the same place where they are injected, without undergoing spatial diffusion. In this scenario the solution of the transport equation can be written as: 
\begin{equation}
n_e(E,z)=\frac{\tau_l(E,z)}{E} \int^{M_\chi}_E dE'\,A\,\frac{dN_e}{dE'}(E')\;\;,\;\;{\rm or\;equivalently,}\;\; \tilde n_e(E)=\frac{\tau_l(E,z)}{E} \int^{M_\chi}_E dE'\,\frac{dN_e}{dE'}(E')\; .
\label{eq:transp1}
\end{equation}
where $\tau_l$ is the time-scale associated to energy losses.
We consider this scenario as our first approximation in Sec.~\ref{sec:bench} (while impacts of possible spatial dependencies of the magnetic field and spatial diffusion of $e^+/e^-$ in the source will be discussed in Secs.~\ref{sec:B} and \ref{sec:diff}).

With these assumptions and disregarding the contribution of substructures, Eq.~(\ref{eq:intgen}) reduces to (see Ref.~\cite{Ullio:2002pj}):
\bea
\nu I_\nu&=&\frac{c\,E_\nu^2}{4\pi\,H_0}\frac{(\sigma_a\,v)}{2}\left(\frac{\Omega_{\chi}\,\rho_c}{M_{\chi}}\right)^2\int dz\,\frac{(1+z)^3\,\Delta^2(z)}{h(z)}\frac{d\tilde N_i}{dE}\,e^{-\tau(z)}\;\;{\rm ~~~~~ for\;annihilating\;DM}\,,\label{eq:Iannsimp}\\
\nu I_\nu&=&\frac{c\,E_\nu^2}{4\pi\,H_0}\frac{\Omega_{\chi}\,\rho_c}{\tau_d\,M_{\chi}}\int dz\,\frac{1}{h(z)}\frac{d\tilde N_i}{dE}\,e^{-\tau(z)}\;\;{\rm ~~~~~ for\;decaying\;DM}\,,\label{eq:Idecsimp}
\eea
with $\rho_c$ being the critical cosmological density.
The quantity $\Delta^2(z)=\langle\delta^2\rangle$ (with $\delta=\rho/\rho_{CDM}$ where $\rho_{CDM}$ is the mean DM density and we will consider $\rho_{CDM}=\bar \rho_\chi=\Omega_{\chi}\,\rho_c$) represents the clumping factor, and encodes all the effects due to the DM clustering.
It is extensively described in, e.g., Ref~\cite{Ullio:2002pj}.
Effects related to the DM halo mass function, the density profile within each halo, and the minimum halo mass in the clumping factor, are discussed  in Sec.~\ref{sec:delta}. 

Plugging Eqs.~(\ref{eq:lumDMa}) and (\ref{eq:lumDMd}) into Eqs.~(\ref{eq:cl1h}) and (\ref{eq:cl2h}), the angular power spectrum of DM can be derived.
Under the same assumptions that lead to Eqs.~(\ref{eq:Iannsimp}) and (\ref{eq:Idecsimp}), one can proceed to analogous simplifications, and again equations found in the prompt gamma--ray emission case (e.g., in Ref.~\cite{Ando:2005xg,ibarra}) can be applied to the radio case by simply replacing $d N_\gamma/dE$ with $d\tilde N_{syn}/dE$.
More precisely, in such a simplified scenario, one finds:
\bea
C_\ell&=&\left(\frac{\sigma\,v}{8\,\pi}\right)^2\left(\frac{\Omega_{\chi}\,\rho_c}{M_{\chi}}\right)^4\,\int\frac{c\,dz}{H(z)\,r(z)^2}\left((1+z)^3\frac{d\tilde N_i}{dE}\right)^2 P_{f_a}(k=\ell/r,z))\,e^{-2\,\tau(z)}\;\;{\rm ~~~ for\;annihilating\;DM}\,,\label{eq:Clannsimp}\\
C_\ell&=&\left(\frac{\Omega_{\chi}\,\rho_c}{4\,\pi\,\tau_d\,M_{\chi}}\right)^2\,\int\frac{c\,dz}{H(z)\,r(z)^2}\left(\frac{d\tilde N_i}{dE}\right)^2 P_{f_d}(k=\ell/r,z)\,e^{-2\,\tau(z)}\;\;{\rm ~~~ for\;decaying\;DM}\,,\label{eq:Cldecsimp}
\eea
where $P_{f_i}$ is the 3D power spectrum of the Fourier transform of $f_i$, where $f_a=\delta^2-\langle\delta^2\rangle$ in the annihilating case, and $f_d=\delta-\langle\delta\rangle$ in the decaying case (thus $P_{f_d}$ is equal to the `standard' non--linear matter power spectrum; for the latter, see, e.g., the review in Ref.~\cite{Cooray:2002dia}). The expression of $P_{f_a}$ can be easily derived adding Eqs.~\ref{eq:cl1h} and \ref{eq:cl2h} (for further details, see Ref.~\cite{Ando:2005xg}) which leads to:
\be
 P_{f_a}(k,z)=P_{f_a}^{1h}+P_{f_a}^{2h}=\int dM \frac{dn}{dM}\,|\tilde{u}(k|M)|^2+\left[\int dM \frac{dn}{dM}\,b(M,z)\,|\tilde{u}(k|M)|\right]^2\cdot P^{lin}(k,z)\;,
\ee
with $\tilde u(k|M)$ being the Fourier transform of $\rho^2(r|M)/\bar \rho_\chi^2$.

\subsubsection{Astrophysical sources}
The radio cosmic background from astrophysical sources is thought to mainly come from two distinct populations, radio loud active galactic nuclei (AGN) and star--forming galaxies (SFG) (for a recent review on radio sources, see, e.g., Ref.~\cite{DeZotti:2009an}). The first dominates source counts at high fluxes, while the latter takes over below $\sim$mJy level. Below mJy level, radio quiet AGNs can also provide an important contribution~\cite{Padovani:2011wj}, which is nevertheless subdominant, and, for simplicity, we will disregard it. 
The luminosity of astrophysical sources is typically written in terms of luminosity at a reference frequency $\nu_0$ where the most robust luminosity function models can be built (i.e., at frequency where more data are present) and including a running to the frequency of interest. We will adopt this method and take $Q=\lum_{\nu_0}$ with $\nu_0=1.4$ GHz and $\nu_0=60\,\mu$m for AGN and SFG, respectively. 

The comoving luminosity function $\phi(\lum,z)=dn/d\,log \lum$ of AGN at 1.4 GHz is taken from Ref. \cite{Massardi:2010sx} where $\phi=n_0/[(\lum/L_*)^a+(\lum/L_*)^b]$ (for values and definition of $n_0$, $a$, $b$, and $L_*$ parameters we refer to their Table 1). The luminosity at other frequencies is obtained assuming a spectrum $\lum\propto\nu^{-\alpha}$.
The model includes two flat--spectrum populations ($\alpha=0.1$) with different evolutionary properties (BL Lacs and Flat Spectrum Radio Quasars) and a steep--spectrum population ($\alpha=0.8$).

Star forming galaxies are typically described by means of their properties in the infrared (IR) band. Although the IR emission may not be generated by the same physical mechanism of radio emission, radio sources are associated to young stars and so tightly correlated to the star formation rate probed at IR frequency. Empirically, the far IR--radio correlation is, at least for high--brightness and low redshift, well established, and follows a nearly linear relation~\cite{DeZotti:2009an} (that we will assume to be redshift independent): $\lum_{1.4\,GHz}=1.16\cdot 10^{-2}L_b/[L_b/\lum_{60\mu m}+(L_b/\lum_{60\mu m})^{3.1}]$ with $L_b=8.8\cdot10^{29}\, {\rm erg s^{-1}Hz^{-1}}$. The luminosity at frequencies other than 1.4 GHz is again obtained through $\lum\propto\nu^{-\alpha}$ with $\alpha=0.75$ (from typical spectral index of optically thin synchrotron sources).
We consider two SFG populations which are expected to be reservoirs of cosmic--rays and so bright in synchrotron emission, starburst galaxies and normal late--type spiral galaxies. Their luminosity function $dn/d\lum$ has been modeled following Ref.~\cite{Saunders:1990kb} with parameters derived from, respectively, `warm' and `cool' IRAS galaxies in Ref.~\cite{Takeuchi:2003za}. 
Our modeling of AGN and SFG agrees well with number counts and angular correlation data.

To compute the angular power spectrum of anisotropies generated by astrophysical sources, we would need to model the spatial profile of luminosity within the source (to get its Fourier transform $u$ in Eqs.~(\ref{eq:cl1h}) and (\ref{eq:cl2h})). This is not as simple as in the DM case (where it follows a universal profile) and in the following we treat astrophysical sources as point sources ($u=1$).
Although this is a crude approximation, it does not affect results up to $\ell\gg10^3$ or $\theta\ll 0.1^\circ$(see, e.g.,Ref.~\cite{Blake:2001bg}, where it is found that multiple components of radio galaxies become important in the correlation function only below few arcmin scales).
With this assumption the form of each of the two terms of the angular power spectrum in 
Eq.~(\ref{eq:cl}) is known, with $C_\ell^{1h}\sim$ const, and the $\ell-$dependence of $C_\ell^{2h}$ set by $P^{lin}$. However, their absolute size depends on $\lum$ and on luminosity integral extrema, so their ratio and the form of total $C_\ell$ is model (and telescope) dependent.

\section{Results}
\label{sec:res}

\begin{figure}[t]
 \begin{minipage}[htb]{0.3\textwidth}
   \centering
   \includegraphics[width=\textwidth]{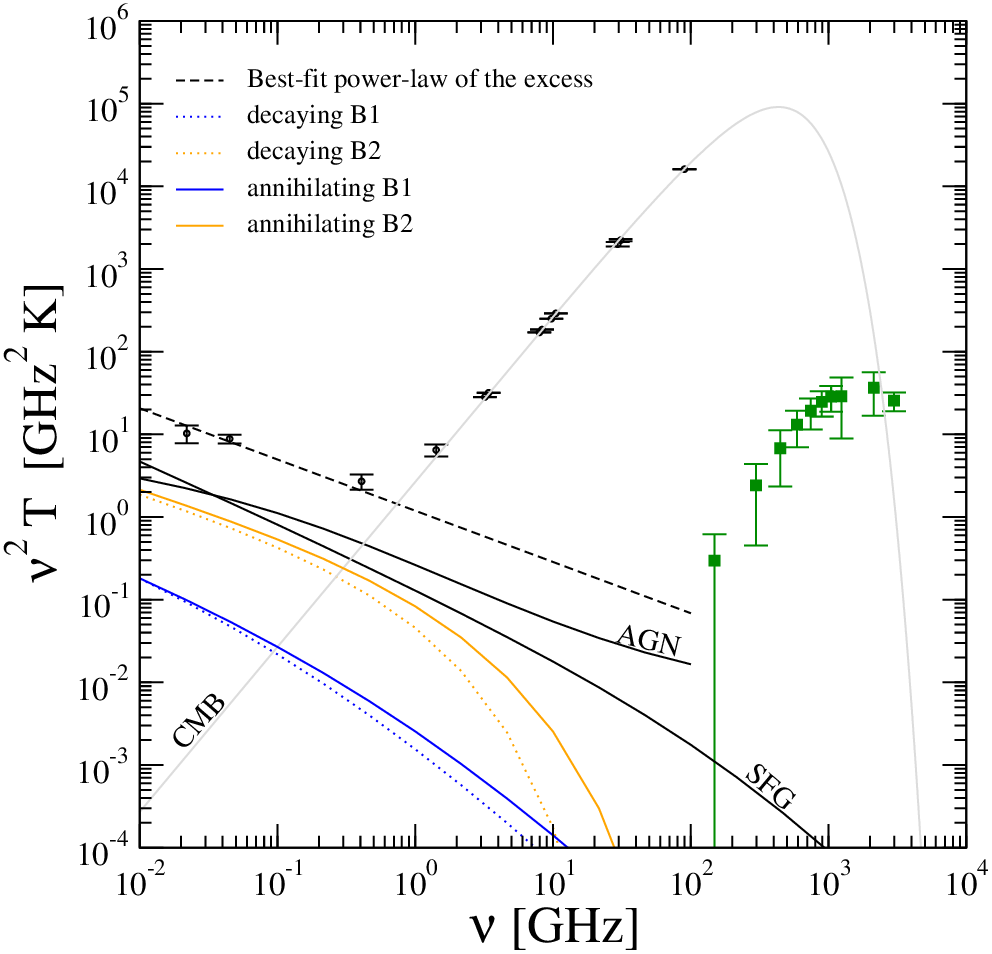}
 \end{minipage}
 \ \hspace{1mm} \
 \begin{minipage}[htb]{0.3\textwidth}
   \centering
   \includegraphics[width=\textwidth]{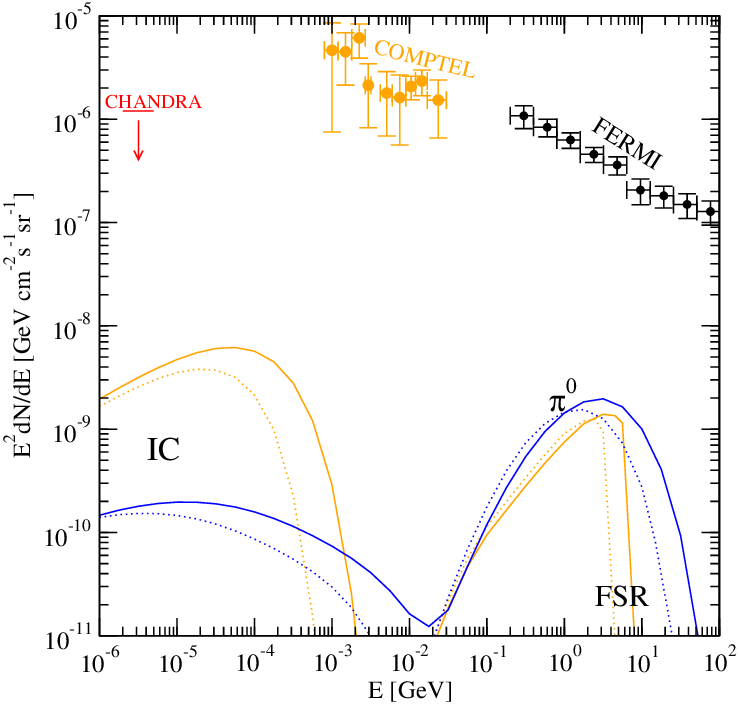}
 \end{minipage}
 \ \hspace{1mm} \
 \begin{minipage}[htb]{0.3\textwidth}
   \centering
   \includegraphics[width=\textwidth]{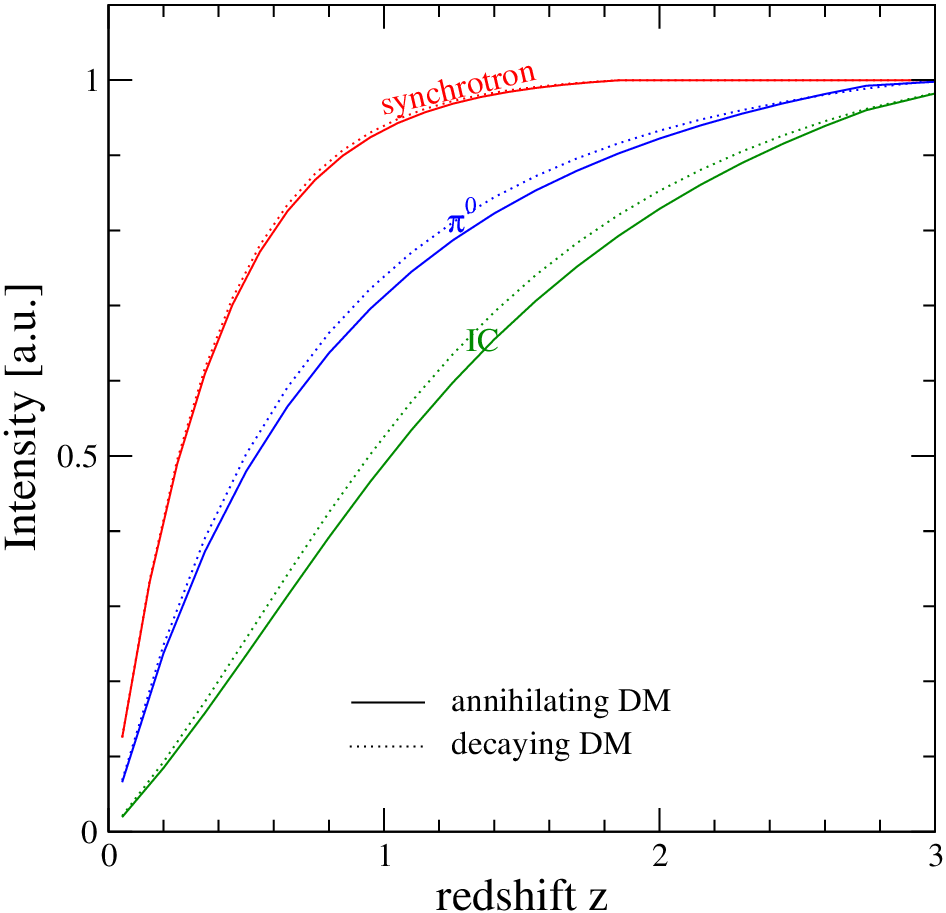}
 \end{minipage}
    \caption{{\bf Total intensity.} Extragalactic radio (left panel) and gamma-ray (central panel) backgrounds for benchmark B1 (blue) and B2 (orange) WIMPs in the annihilating (solid) and decaying (dotted) cases (see text and Table~\ref{tab:bench} for details). For the radio case, main astrophysical source contributions (black--solid), CMB (black--dotted), and the best--fit of ARCADE data (black--dashed) are also shown~\cite{Seiffert:2009xs}. The right panel shows contributions at different redshifts (i.e., $I^{-1}\int_0^z dz'\,dI/dz'$) for emissions given by synchrotron radiation at 1 GHz, inverse Compton on CMB at 1 MeV, and $\pi^0$-decay at 1 GeV, in the benchmark case B1 (all normalized to unity).}
\label{fig:totI}
 \end{figure}

\subsection{Benchmark cases}
\label{sec:bench}
 
\begin{table}[t]
\begin{center}
\begin{tabular}{|c|c|c|c|c|}
\hline
Name&Mass& $(\sigma_a v)\,[{\rm cm^3s^{-1}}]$& $\tau$ [s] & Dominant\\
 & [GeV] & annihilating case & decaying case&final state   
\tabularnewline
\hline
\hline
B1& 100 &$3\cdot10^{-26}$& $4\cdot10^{28}$ & $b-\bar b$
\tabularnewline
\hline
\hline
B2& 10 &$3\cdot10^{-26}$& $5\cdot 10^{27}$ & $\mu^+-\mu^-$
\tabularnewline
\hline
\end{tabular}
\end{center}
\caption{Benchmark DM models.}
\label{tab:bench}
\end{table}

In order to have a feeling about the expected DM signals and how they compare to astrophysical emissions, we first show results for few benchmark cases.
We consider a neutralino--like DM candidate (model B1) with mass $M_{\chi}=100$ GeV, `thermal' annihilation rate $(\sigma_a v)=3\cdot10^{-26}\, {\rm cm^3s^{-1}}$ and final state of annihilation into quarks $b-\bar b$ (and similarly a decaying case with $\tau=4\cdot10^{28}$ s). Motivated by recent claims from direct detection experiments and by a possible DM interpretation of the ARCADE excess~\cite{Fornengo:2011cn}, we introduce also a DM candidate (model B2) with mass $M_{\chi}=10$ GeV, again with  `thermal' annihilation rate, but annihilating into the leptonic final state $\mu^+-\mu^-$ (to enhance the synchrotron signal), and an analogous decaying case with $\tau=5\cdot 10^{27}$ s.

We adopt the halo mass function from Ref.~\cite{Sheth:1999mn}, the concentration of halos from
Ref.~\cite{MunozCuartas:2010ig}, and the DM distribution in halos following a NFW profile~\cite{Navarro:1996gj}. No substructures are considered in this benchmark scenario and the minimum halo mass is set to $M_{cut}= 10^5 M_{\odot}$. The magnetic field is assumed to be constant in space and time with $B=10\,\mu G$, and $e^+/e^-$ are assumed to radiate at the same place where they are injected. Each of these assumptions will be discussed and relaxed in the next Sections.

Throughout the paper, we will stick to a 6--parameter $\Lambda$CDM model derived by the WMAP team combining CMB data with Baryon--Acoustic--Oscillation (BAO) and $H_0$ measurements~\cite{Komatsu:2010fb}\footnote{Note, however, that even our relatively very little ignorance about cosmological parameters can induce an $\mathcal{O}(1)$ uncertainty in some results, since, e.g., terms like $\Omega_{CDM}$ and $\sigma_8$ enters to the fourth power in $C_\ell$.}. Parameters relevant for our purposes are $\Omega_m=0.27$, $\Omega_\Lambda=0.73$, $\Omega_{CDM}=0.23$, $h=0.70$, $n_s=0.96$, and $\sigma_8=0.8$.

In Fig.~\ref{fig:totI}a, we plot the total intensities.
Note that the DM contribution in those benchmark scenarios is roughly at the same level of astrophysical emissions.
The extra--galactic sky temperature reported by the ARCADE 2 collaboration~\cite{Fixsen:2009xn} cannot be explained by ordinary models of AGN + SFG.
In Ref.~\cite{Fornengo:2011cn}, we discussed the possibility of fitting ARCADE data in terms of a DM--induced synchrotron emission. WIMP models can actually account for the excess, with viable scenarios being only slightly more optimistic than the benchmark cases considered here~\cite{Fornengo:2011cn}.
Comparing Fig.~\ref{fig:totI}a with Fig.~\ref{fig:totI}b, one can see that radio extragalactic data are more constraining than the $\gamma$-ray counterpart for WIMP models with dominant annihilation/decay channel into leptons (by roughly one order of magnitude in the B2 case); on the contrary, for WIMP models with significant $\pi^0$ production (as in the case of dominant hadronic final states B1), gamma and radio data have roughly the same constraining power. 

We have been considering energy losses due to synchrotron radiation and to IC scattering on CMB only. The effect of the electron/positron interactions with the interstellar radiation field (ISRF) is difficult to model within this formalism since it is mostly related to the properties of the astrophysical sources hosted within each halo, rather than to the halo mass. However, the IC loss associated to the ISRF is expected to be generically subdominant in the computation of extragalactic signals. To understand this point, we included energy losses on a simple ISRF with constant density of $1\,{\rm ev\,cm^{-3}}$ in all halos at any $M$ and $z$. This density is roughly the mean optical ISRF density of the Milky Way and can be considered as an overestimate of the actual field since it is known that either lager (e.g., cluster) and smaller (e.g., dwarf spheroidal galaxies) structures have much lower density. The effect on the total intensity curves in Fig.~\ref{fig:totI}a is an overall depletion of $\sim15\%$. We therefore assume energy losses due to IC on ISRF to be subdominant, and we disregard them in the rest of the paper.

In Fig.~\ref{fig:totI}c, the redshift distribution of main radio, X-, and $\gamma$-ray signals is shown.
In the case of inverse Compton scattering on CMB, it is completely set by the evolution of the DM spatial distribution. Indeed, the dilution of CMB energy density due to the expansion of the Universe affects both energy losses and IC power, having basically no net effects on the emission. There can be a further suppression in the emission at low $z$ if synchrotron losses take over at late times. Moreover, the location of the IC peak is roughly redshift independent (and so the energy of the emitted electrons), since the factor $1/(1+z)$ arising from emission to the observer is exactly compensated by the increase $(1+z)$ in the CMB target photon energy~\cite{Profumo:2009uf}.
For $\pi^0$-decay emission, there is an extra decrease with redshift which depends on the WIMP annihilation/decay spectrum, since the emission is given by $dN_\gamma/dE ((1+z)E)$. Accordingly, the contribution at high redshifts becomes less significant with respect to the IC on CMB case.

An important conclusion that can be drawn from Fig.~\ref{fig:totI}c is that the synchrotron total intensity is provided by emissions at low redshifts, lower than in the case
of the IC case. This is easy to understand and is due to the rapid increase in the energy loss associated to IC scattering on CMB (scaling as $(1+z)^4$), while a significant increase with $z$ of the magnetic field is not expected.

\begin{figure}[t]
 \begin{minipage}[htb]{0.3\textwidth}
   \centering
   \includegraphics[width=\textwidth]{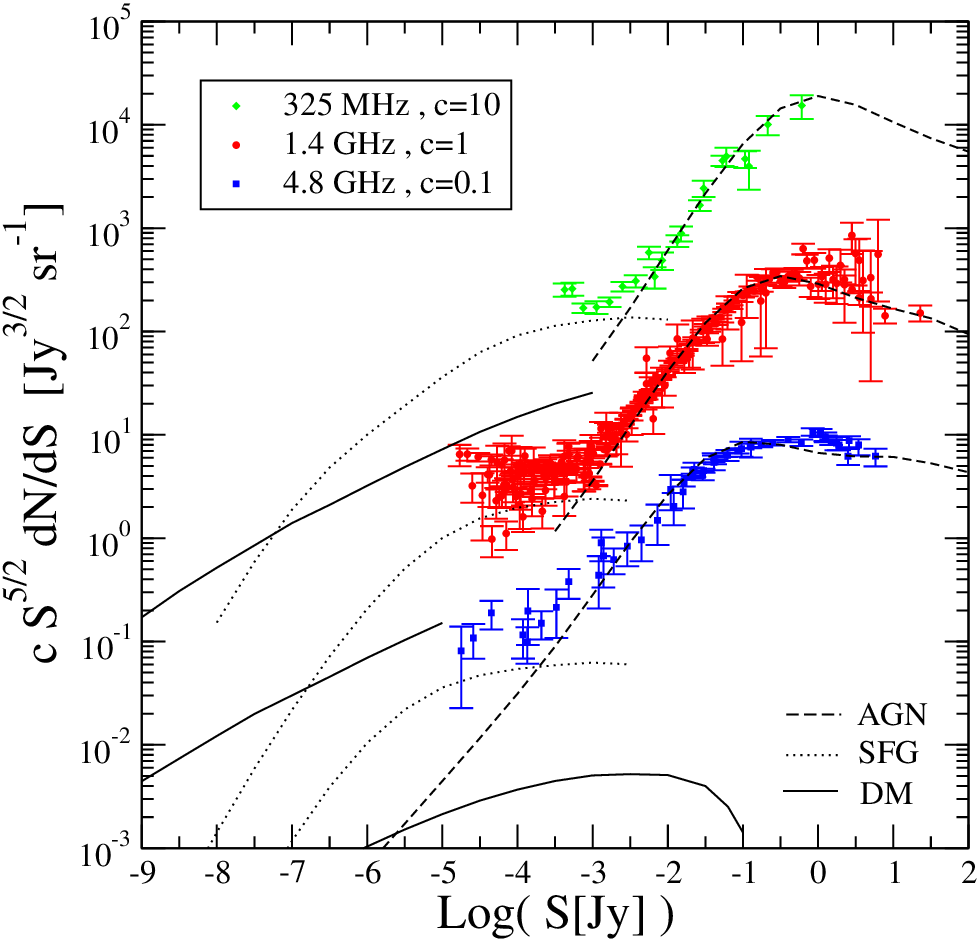}
 \end{minipage}
 \ \hspace{2mm} \
 \begin{minipage}[htb]{0.3\textwidth}
   \centering
   \includegraphics[width=\textwidth]{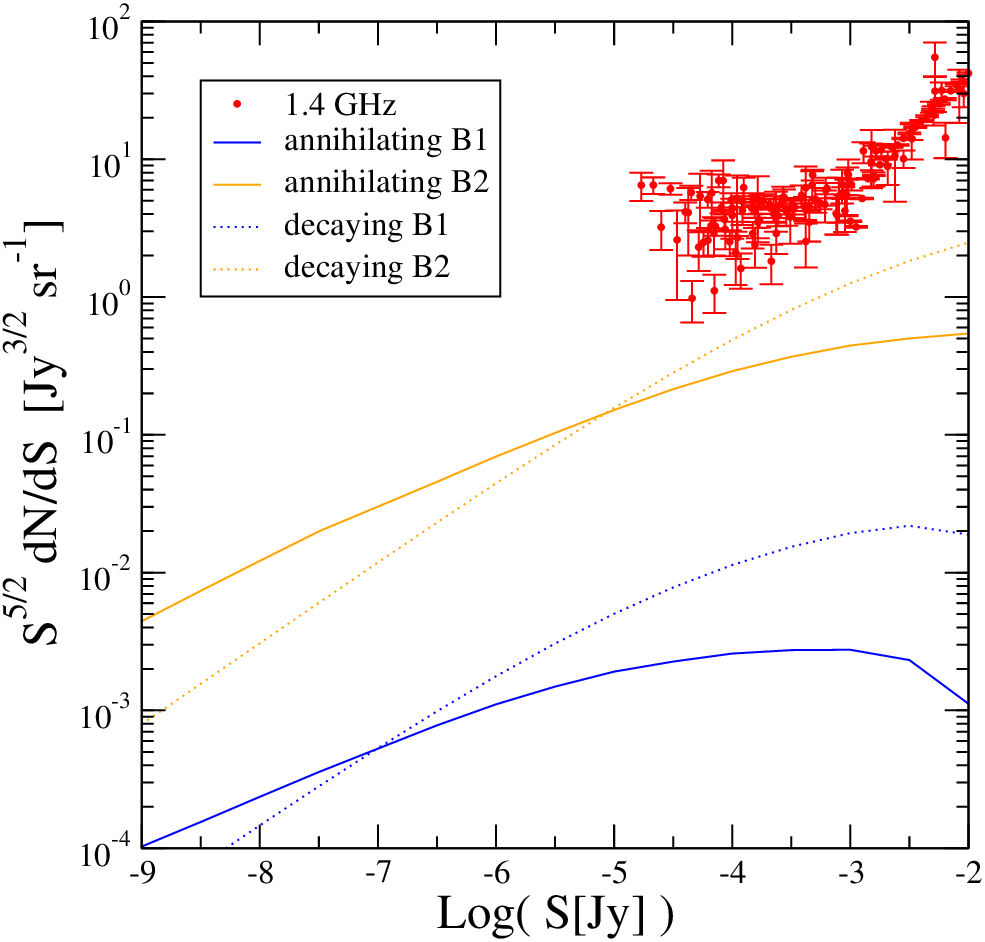}
 \end{minipage}
 \ \hspace{1mm} \
 \begin{minipage}[htb]{0.3\textwidth}
   \centering
   \includegraphics[width=\textwidth]{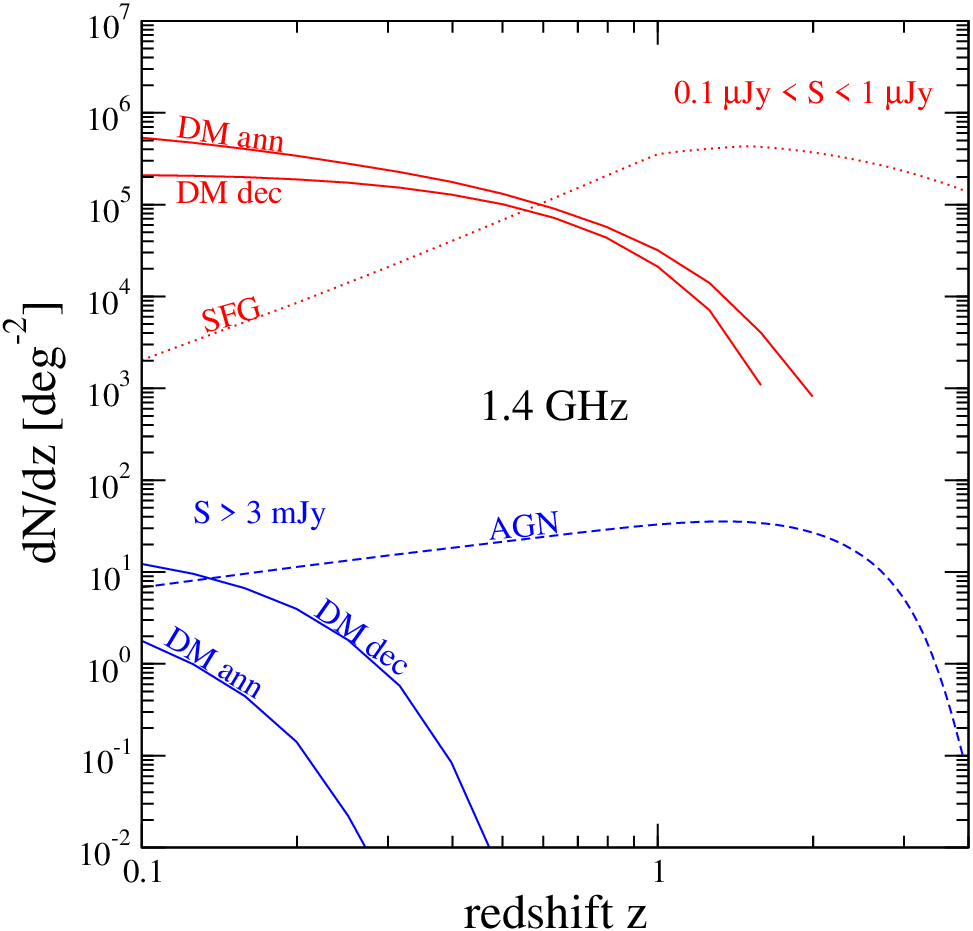}
 \end{minipage}
    \caption{{\bf Source counts.} {\it Left:} Differential number counts for AGN (dashed), 
    star--forming galaxies (dotted), and the benchmark B2 annihilating DM model (solid) described in the text. Data are from Ref.~\cite{DeZotti:2009an} and are multiplied by a 
    factor $c$ (given in the figure inset) for clarity. {\it Central:} Differential number counts at 1.4 GHz for all benchmark DM models described in Table~\ref{tab:bench}. {\it Right:} Redshift distribution of a bright ($S>3$ mJy, lower) and a faint ($0.1\,\mu {\rm Jy}<S<1\, \mu {\rm Jy}$, upper) samples for astrophysical sources (only dominant population is shown) and the benchmark DM models B2.}
\label{fig:counts}
 \end{figure}

In Fig.~\ref{fig:counts} number counts are shown.
The annihilating WIMP model B2, which leads to a total intensity comparable to astrophysical contributions (as shown in Fig.~\ref{fig:totI}), is compared to AGN and SFG counts in Fig.~\ref{fig:counts}a. AGNs largely dominate at strong brightness, while SFGs take over below 
mJy--fluxes. The DM contribution becomes dominant in the sub-$\mu$Jy regime. The corresponding decaying case has a steeper spectrum (see Fig.~\ref{fig:counts}b), and so, although at large brightness it is relatively more important than the annihilating case (but still subdominant), it takes over only well below the $\mu$Jy level.
This is because of the $\rho^2$ scaling in the annihilating case (combined with the growing of the concentration parameter as the halo mass decreases) which makes the smaller and fainter structures relatively more important than large and bright halos.

Fig.~\ref{fig:counts}c shows again (along the same line of Fig.~\ref{fig:totI}) that most of the WIMP--induced radio emission comes from sources at very low redshift.
For large brightness, AGN dominates and the DM contribution is subdominant basically at all redshift. On the other hand, in the sub-$\mu$Jy regime, DM sources are more numerous than SFG up to $z\sim 1$.
The precise shape of SFG counts is not completely understood yet, and curves in Fig.~\ref{fig:counts} represent one possible model. However, the overall normalization and the fact that SFGs peak at $z\gtrsim1$ are rather robust predictions~\cite{DeZotti:2009an}.
It means that we can firmly conclude that a WIMP source population which provides a substantial contribution to the total intensity has to dominate in the source count at low brightness ($S\lesssim \mu$Jy) and low redshift ($z\lesssim 1$).

\begin{figure}[t]
 \begin{minipage}[htb]{0.3\textwidth}
   \centering
   \includegraphics[width=\textwidth]{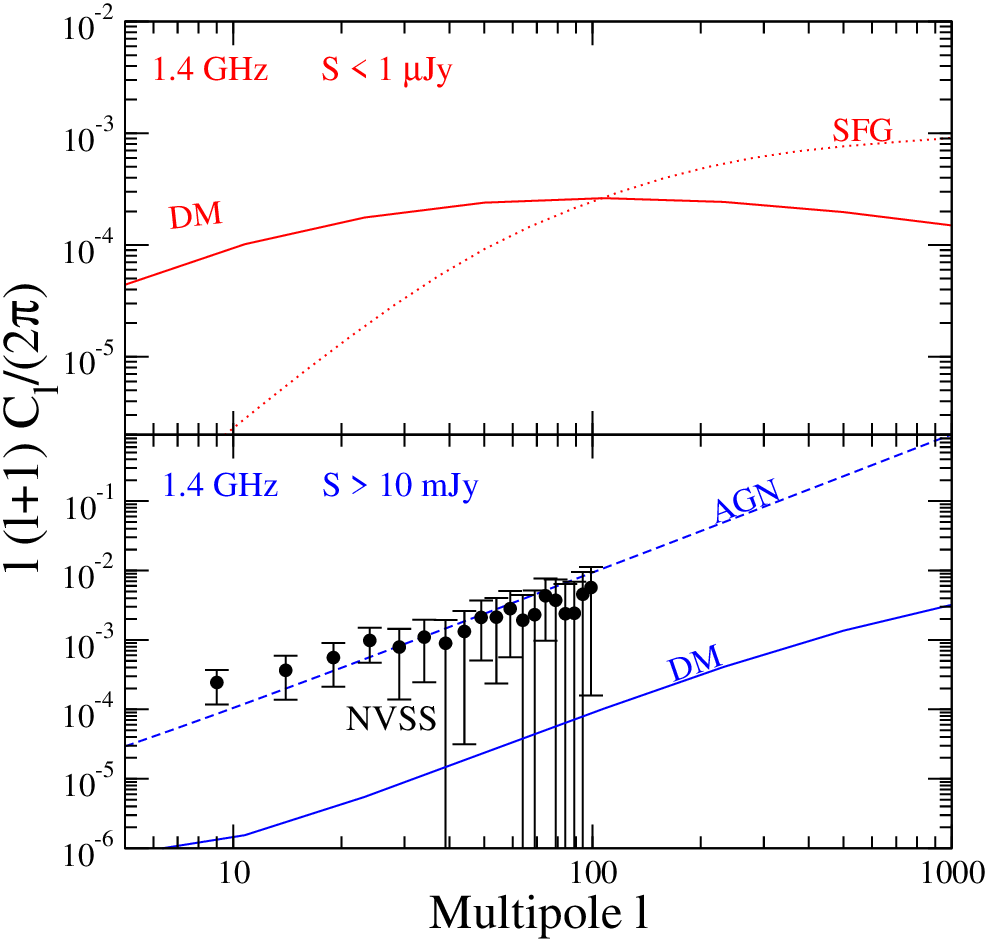}
 \end{minipage}
 \ \hspace{1mm} \
 \begin{minipage}[htb]{0.3\textwidth}
   \centering
   \includegraphics[width=\textwidth]{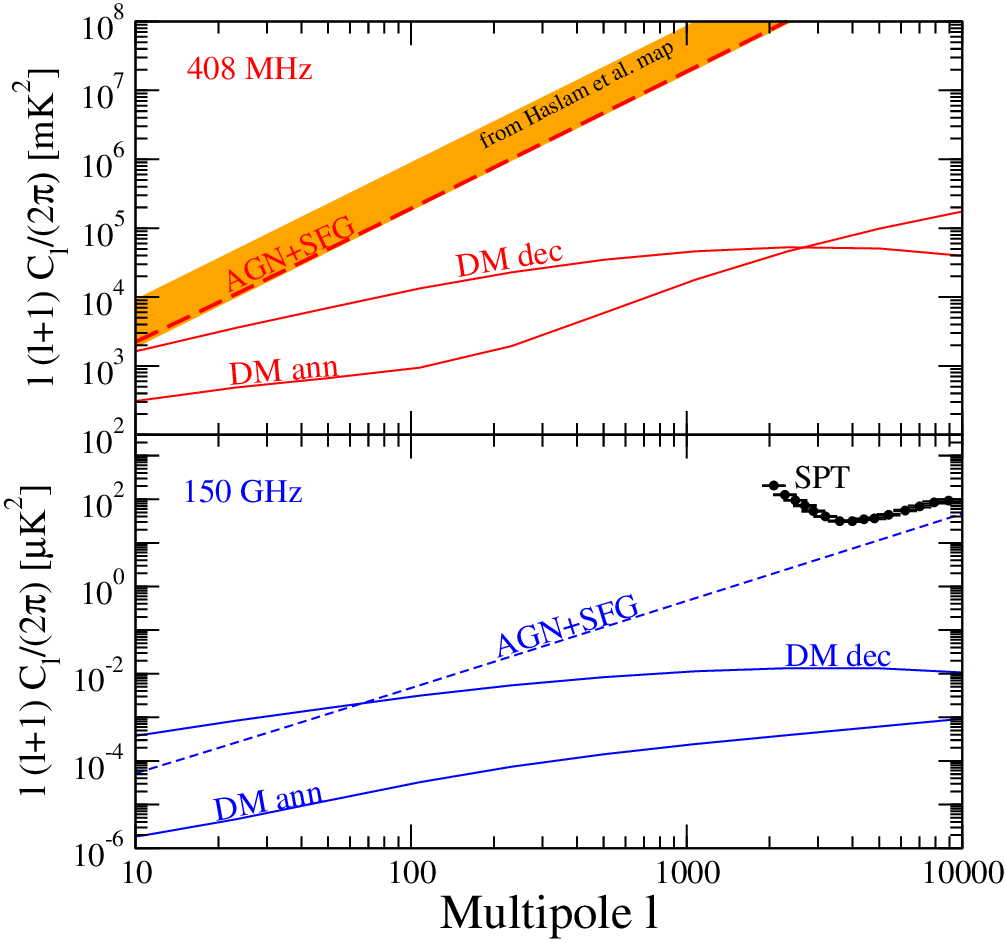}
 \end{minipage}
 \ \hspace{1mm} \
 \begin{minipage}[htb]{0.3\textwidth}
   \centering
   \includegraphics[width=\textwidth]{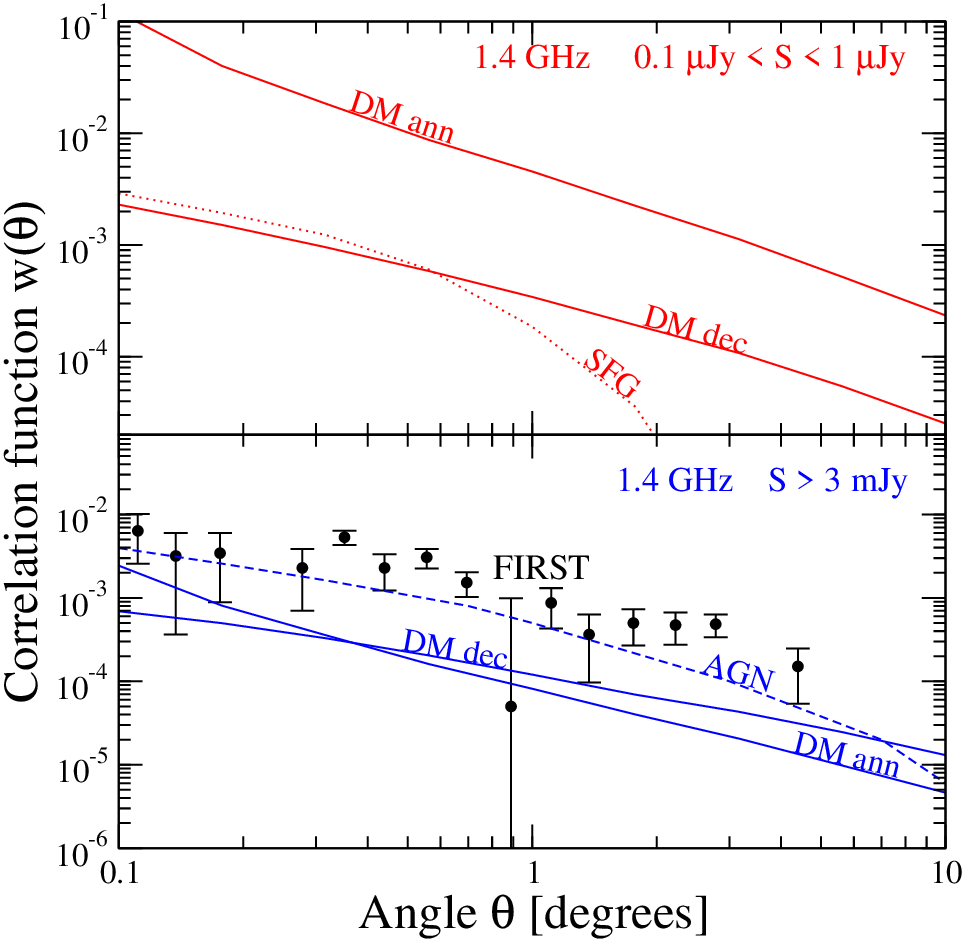}
 \end{minipage}
    \caption{{\bf Angular correlation.} {\it Left:} Angular power spectrum at 1.4 GHz of a bright ($S>10$ mJy, lower) and a faint ($S<1\, \mu {\rm Jy}$, upper) samples for astrophysical sources (only the dominant population is shown) and the benchmark B2 annihilating DM model. Data points are derived from NVSS in Ref.~\cite{Blake:2004dg}. {\it Central:} Angular power spectrum computed including contributions from sources at all brightness at 408 MHz (upper, observational band derived from Ref.~\cite{Haslam}, see Refs.~\cite{Regis:2011ji,LaPorta:2008ag}) and 150 GHz (lower, data from Ref.~\cite{Reichardt:2011yv}) for astrophysical sources, the benchmark DM models B2 (at 408 MHz), and a WIMP model (at 150 GHz) with $M_{\chi}=1$ TeV and $(\sigma_a v)=3\cdot10^{-23}\, {\rm cm^3s^{-1}}$ (annihilating) or $\tau=10^{26}$ s (decaying) (i.e., fitting the PAMELA positron excess~\cite{Adriani:2008zr}).  {\it Right:} Angular correlation function at 1.4 GHz of a bright ($S>3$ mJy, lower) and a faint ($0.1\,\mu {\rm Jy}<S<1\, \mu {\rm Jy}$, upper) samples for astrophysical (only dominant population is shown) and DM benchmark models B2. Data are from~\cite{Overzier:2003kg}.}
\label{fig:cl}
 \end{figure}

In Fig.~\ref{fig:cl}, we show the angular correlation of radio sources. Dimensionless $C_\ell$ and $w$ are normalized to, respectively, the mean intensity $\langle I \rangle$ and  number counts $N$ corresponding to the astrophysical contributions.
For what concerns the power spectrum, the one--halo term of Eq.~(\ref{eq:cl1h}) (which is basically a Poisson noise term, $C_\ell\sim$ const, up to very large multipoles) dominates at large brightness (namely, at brightness currently covered by data), where the DM contribution is again much smaller than the AGN one.
In the sub--$\mu$Jy regime, the two--halo term is instead the most important, and the DM population starts dominating, in particular at low multipoles, i.e., large scales.
Note also that the DM power spectrum is completely different from a flat $C_\ell$ in both Figs.~\ref{fig:cl}a and \ref{fig:cl}b (except when taking only brightest halos), telling us that it is mostly provided by the two--halo term. This is again due to the fact that the DM source peaks at very low redshift, which explains also why the DM tends to have more power on large scales.
For data points corresponding to observations which integrate over all brightness, like the Haslam et al. map~\cite{Haslam} at 408 MHz and the SPT survey~\cite{Reichardt:2011yv} at 150 GHz, the $C_\ell$ are mostly given by the brightest objects and the DM contribution is only marginally relevant at $\ell\lesssim100$ (where however any estimate of the extragalactic spectrum is undermined by uncertainties in the Galactic anisotropies).
The decaying case is again less (more) favorable for faint (bright) fluxes with respect to the annihilating DM. Note also that it has relatively more power on large scales, stemming from the $\rho^2$ dependence of the signal in the annihilating case which enhances the relevance of small scales.
A similar discussion applies also for the angular correlation function plotted in Fig.~\ref{fig:cl}c, with slightly more favorable conclusion for WIMPs.

A big caveat in our analysis is that we have been treating the source population given by DM induced signals as a separate population with respect to AGN and SFG.
On the other hand, any astrophysical source is embedded in a (typically much fainter) DM halo. This means that if one subtracts bright sources to isolate, say, sub--$\mu$Jy sources, he might be subtracting also emissions from the corresponding DM halos, so possibly a significant fraction of the total DM contribution.
Therefore plots in Figs.~\ref{fig:counts} and \ref{fig:cl} are theoretical predictions, but they could be difficult to be fully tested through observations, since the latter will be probably biased towards DM sources hosting a faint baryonic counterpart.
A possible way to circumvent this issue is to compute the total emission
for each structure in the universe (namely to add the DM and astrophysical signals arising in the same source), 
and then comparing the results between the cases with/without DM (and applying different brightness cuts).
This would require to associate to any DM halo of mass M and redshift z
all the different possible astrophysical source population and spatial distribution in a statistical way.
This is theoretically very challenging and deserves a dedicated analysis.

\subsection{Discussion on assumptions}
\label{sec:disc}
\subsubsection{Dark matter clustering}
\label{sec:delta}

As described in Sec.~\ref{sec:thmod}, we consider a `semi-analytical' procedure where the contribution from halos of all masses is integrated by using analytical functional forms, which are tuned to reproduce results from numerical N--body simulations.
The most important ingredients are the halo mass function, concentration, and density profile, the minimum halo mass, and the subhalo abundance.

The halo mass function is taken from Ref.~\cite{Sheth:1999mn}. Considering more recent and accurate results from numerical simulations, the agreement is within 10--20\% (see, e.g., Fig.~7 in Ref.~\cite{Tinker:2008ff}). We choose the function from Ref.~\cite{Sheth:1999mn} since it is well behaved at all masses, while results from, e.g., Ref.~\cite{Tinker:2008ff} would need some arbitrary extrapolation at small masses.

\begin{figure}[t]
 \begin{minipage}[htb]{0.3\textwidth}
   \centering
   \includegraphics[width=\textwidth]{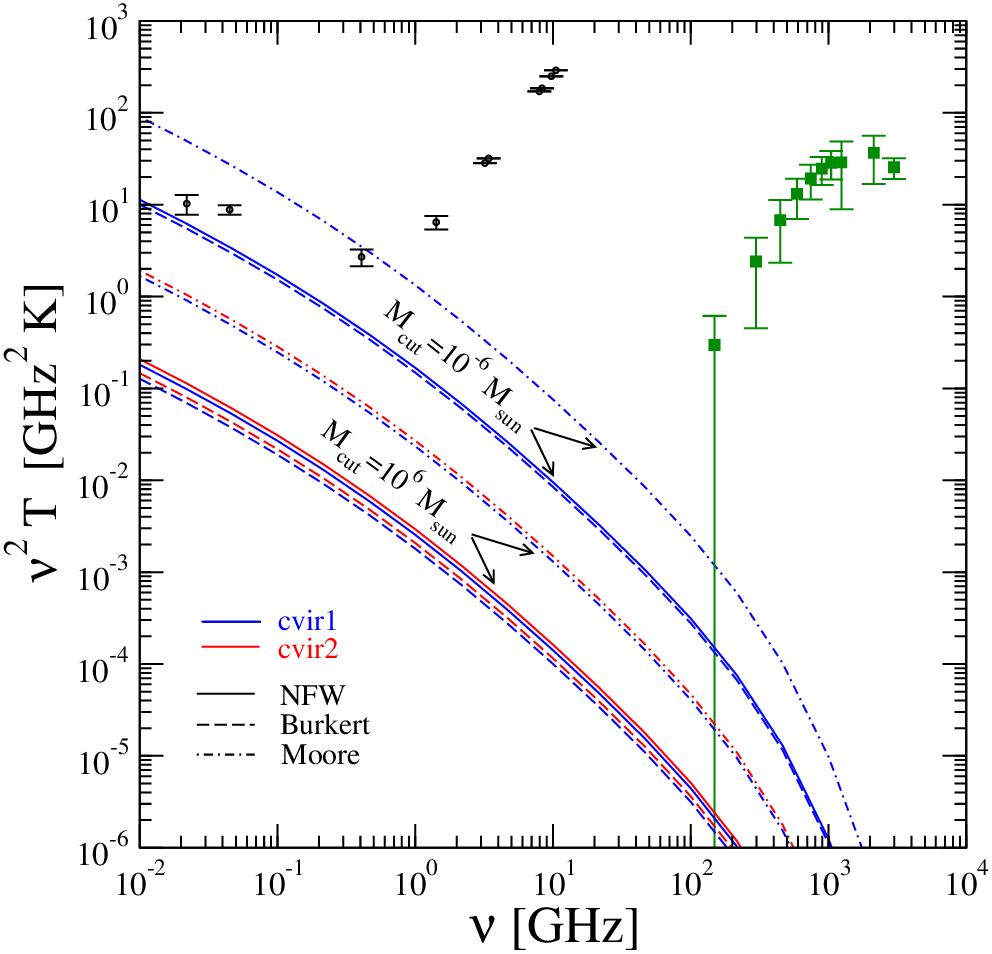}
 \end{minipage}
 \ \hspace{1mm} \
 \begin{minipage}[htb]{0.3\textwidth}
   \centering
   \includegraphics[width=\textwidth]{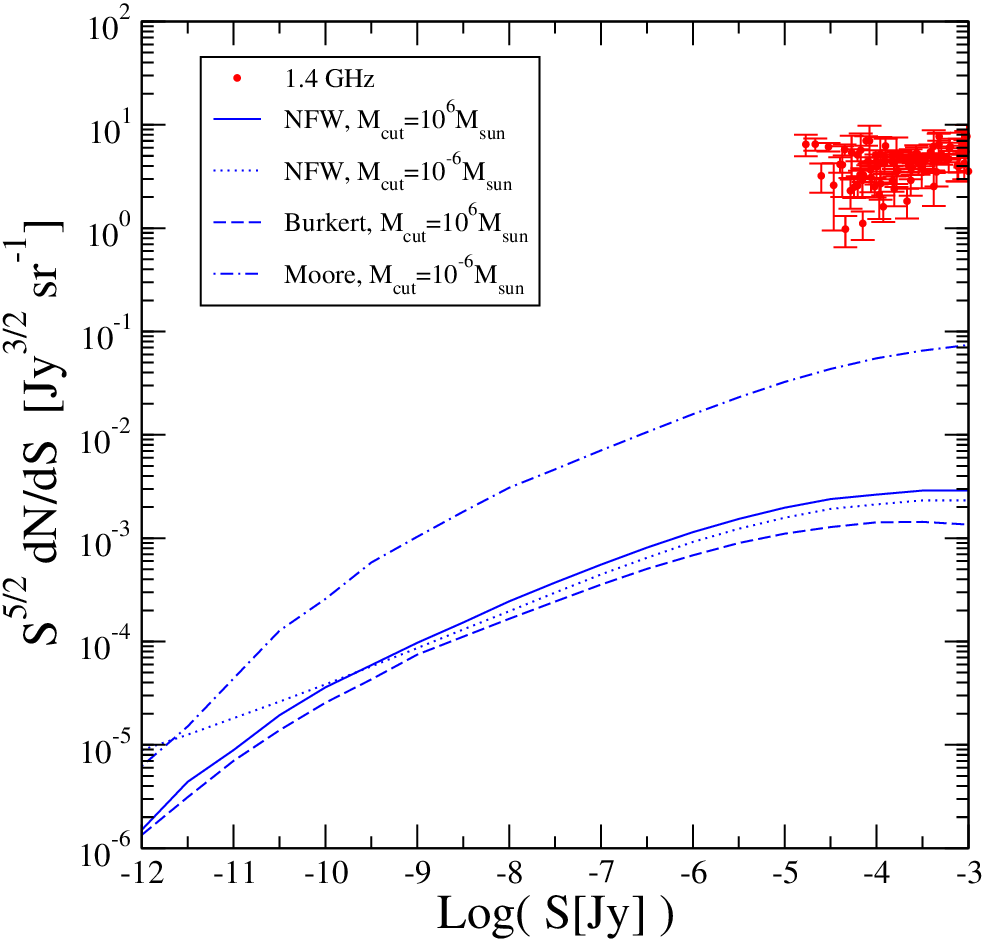}
 \end{minipage}
 \ \hspace{1mm} \
 \begin{minipage}[htb]{0.3\textwidth}
   \centering
   \includegraphics[width=\textwidth]{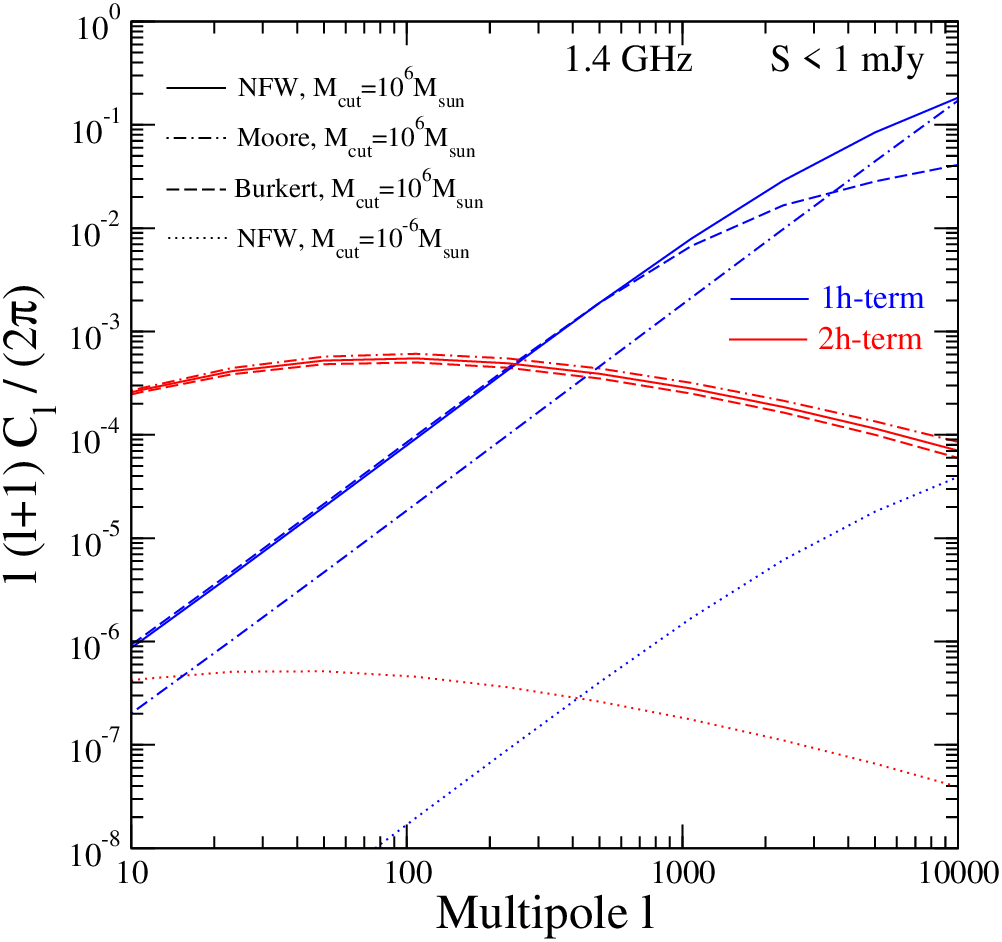}
 \end{minipage}
    \caption{{\bf Clustering.} {\it Left}: Total intensity for the benchmark annihilating model B1 considering three different halo profile, NFW (solid), Burkert (dashed), and Moore (dashed-dotted), two different minimum halo mass, $M_{cut}= 10^6 M_{\odot}$ (lower) and $M_{cut}= 10^{-6} M_{\odot}$ (upper), and two different models for the concentration-mass relation, from~\cite{Prada:2011jf} (red) and \cite{MunozCuartas:2010ig} (blue). {\it Central}: Number counts at 1.4 GHz for the benchmark annihilating model B1 considering an NFW profile with $M_{cut}= 10^6 M_{\odot}$ (solid) and $M_{cut}= 10^{-6} M_{\odot}$ (dotted), together with Burkert (dashed), and Moore (dashed-dotted) profiles (with $M_{cut}= 10^6 M_{\odot}$). {\it Right:} Angular power spectrum for the same models of central panel. The one (blue) and two (red) halo terms are shown separately. }
\label{fig:DMdistr}
 \end{figure} 

We consider Refs.~\cite{Prada:2011jf} and \cite{MunozCuartas:2010ig} for most recent results regarding the concentration--mass relation in the CDM scenario.
The main difference (and most interesting feature) of Ref.~\cite{Prada:2011jf} with respect to Ref. \cite{MunozCuartas:2010ig} is that they find a flattening and upturn with increasing mass for halo concentrations at high redshifts. However, this is not particularly relevant for our purposes since the synchrotron signal is mostly generated at low redshift and in low mass halos, where they are in good agreement. Indeed in terms of radio emission, the two models are basically coincident as shown in Fig.~\ref{fig:DMdistr}. For convenience we choose the halo concentration in Ref.~\cite{MunozCuartas:2010ig} as our benchmark parametrization since it is defined in terms of $c_{vir}-M_{vir}$ (we take into account a scatter in this relation by assuming that $c_{vir}$ follows a log--normal distribution with a width of 20\%~\cite{Ullio:2002pj}).
Taking an observationally--driven point of view, one could use the relation between the halo core radius $r_0$ and central density $\rho_0$ found in Ref.~\cite{Donato:2009ab} for cored DM profiles, and express it in terms of $c_{vir}(M_{vir})$. We found that the relation $r_0\rho_0\simeq {\rm const}\simeq2$~\cite{Donato:2009ab} implies halo concentrations which are similar to the results of simulations for $M=10^6-10^8 M_{\odot}$, while significantly lower concentrations are obtained for larger masses. This relation stems from measurements of nearby galaxies and so we do not attempt to use it in our computations which involve both larger and smaller structures and halos at $z>0$. However, even extrapolating somehow the relation in those ranges, we do not expect dramatically different predictions, since the DM signal is mainly given by the contribution of the smallest halos.

A standard reference for the concentration--mass relation adopted in estimates of extragalactic signals has been Ref.~\cite{Bullock:1999he}.
Now it is found to be not consistent with most recent simulations~\cite{MunozCuartas:2010ig,Prada:2011jf}, mainly because underpredicts the concentrations for more massive halos.
In our computation, however, it leads to, at most, $\mathcal{O}(1)$ difference (depending on the choiche of $K$ and $F$ parameters in the model), so it does not represent a large source of difference with respect to past works. 

The main uncertainty comes from the extrapolation at low masses, where unfortunately we don't have much simulation data.
In this case, different models can lead to significantly different results. To bracket such uncertainty we consider a naive extrapolation of the model of Ref.~\cite{MunozCuartas:2010ig} down to $10^{-6} M_{\odot}$ and a case where there is no contribution below $10^6 M_{\odot}$.
As shown in Fig.~\ref{fig:DMdistr}a, for the intensity there is a factor of 50 between this two pictures.
Such uncertainty is even larger than the one associated to the halo density profile, whose effect is shown by plotting three different cases, NFW, Burkert and Moore.
Only the latter, which is a quite extreme choice, provides a significant boost.
Note that only the NFW profile is strictly consistent with simulations~\cite{MunozCuartas:2010ig,Prada:2011jf,Bullock:1999he} adopted for halo concentration. When considering Burkert and Moore profiles, we are somehow implicitly assuming that they arise from an evolution of an NFW profile induced by `non-cosmological' effects associated to barionic physics. 
It's interesting to note that, for number counts, the choice of $M_{cut}$ makes, in practice, little difference. Indeed, very low mass halos have very low luminosities and therefore significantly increase the counts only in a range very far from experimental sensitivities. At larger luminosities the small difference simply comes from the normalization of $dn/dM$ which is given by $\int_{M_{cut}} dM\,dn/dM\,M=\Omega_m\rho_c$.
Since we stop at a given halo mass $M_{cut}$, the counts stop at brighter luminosities for steeper profiles, and the faint end of number counts in the Moore case drops at larger fluxes with respect to NFW and Burkert cases.
A possible argument in favor of $M_{cut}\gtrsim 10^6 M_{\odot}$ relies on the need of a reasonable amount of barionic matter in order to generate a magnetic field significantly stronger than the cosmological one (which is $\sim nG$). The fact that smallest DM halos would not (presumably) host a large amount of luminous mass component implies that synchrotron radiation would be accordingly much smaller than in more `standard' halos. This will be discussed in Sections \ref{sec:B} and \ref{sec:diff}.

Fig.~\ref{fig:DMdistr}c shows the impact of halo density profile and minimum halo mass on the angular power spectrum.
As for number counts, the choice of $M_{cut}$ makes little difference for the dimension--full $C_\ell$ (except for mildly enhancing the relative importance of the $2h$ term with respect to $1h$), since the correlation of faintest halos becomes important only if focusing on emissions at very low brightness. Dimensionless $C_\ell$ are strongly rescaled due to the much higher factor $\langle I \rangle$ (reflecting what found in Fig.~\ref{fig:DMdistr}a) in the normalization.
For the 1--halo term, the Burkert and Moore profiles show, respectively, less and more power at small scales ($\ell \gtrsim 10^3$) with respect to the NFW case, as expected. The 2--halo term is instead mostly set by the linear power spectrum of Eq.~(\ref{eq:cl2h}) (which is obviously the same for all the three cases).
Note also, as already mentioned above, that the 2--halo term dominates up to large multipoles.   
Fig.~\ref{fig:DMdistr}c is obtained by integrating basically over all brightnesses ($S<1$ mJy). 
When cutting contributions from high--flux sources, the 1--halo term decreases much more rapidly than the 
2--halo term, which therefore can dominate the whole spectrum up to very large multipoles.

Summarizing, the total uncertainty from DM clustering properties (without introducing substructures) in the intensity, number counts, and angular power spectrum of annihilating DM reaches about two orders of magnitude.
Note from Eqs.~(\ref{eq:Idecsimp}) and (\ref{eq:Cldecsimp}) that, under the approximation of spatially constant magnetic field and diffusion 
which do not vary with the halo mass, the results in the decaying case are completely independent from all the uncertainties discussed in this Section.

\begin{figure}[t]
 \begin{minipage}[htb]{0.3\textwidth}
   \centering
   \includegraphics[width=\textwidth]{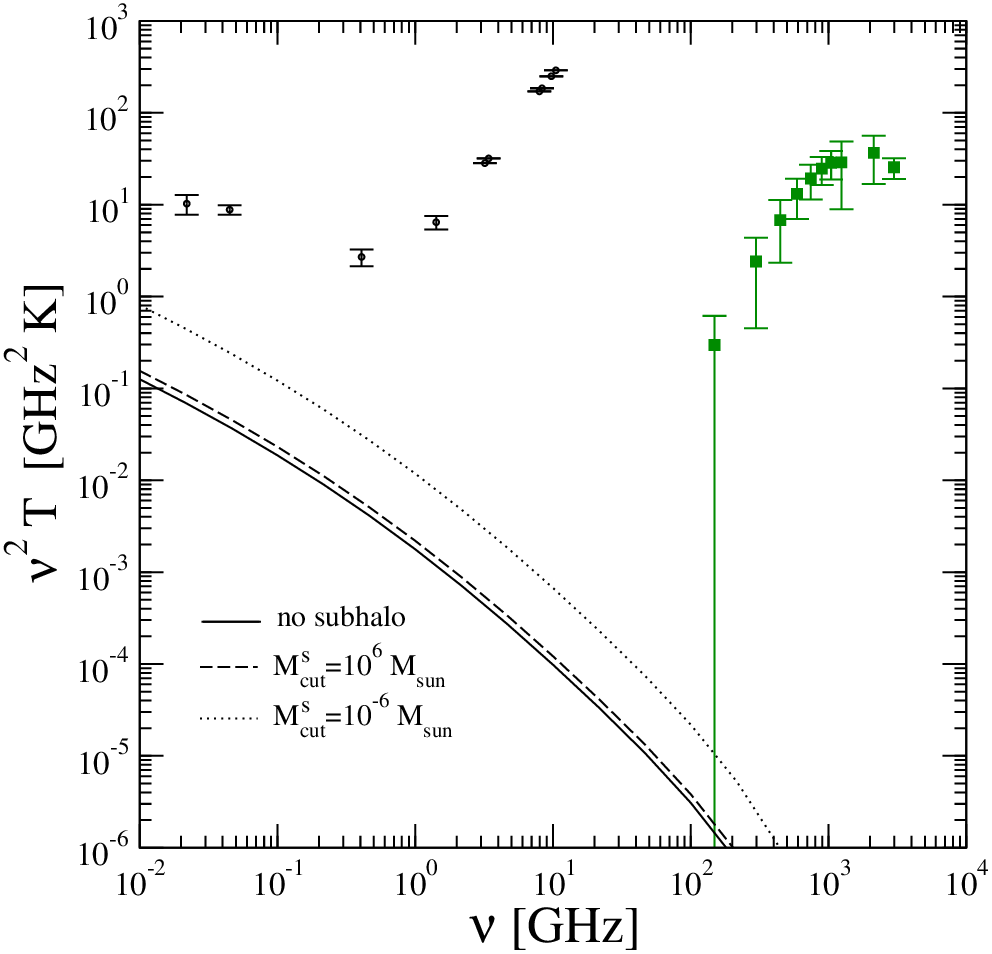}
 \end{minipage}
 \ \hspace{1mm} \
 \begin{minipage}[htb]{0.3\textwidth}
   \centering
   \includegraphics[width=\textwidth]{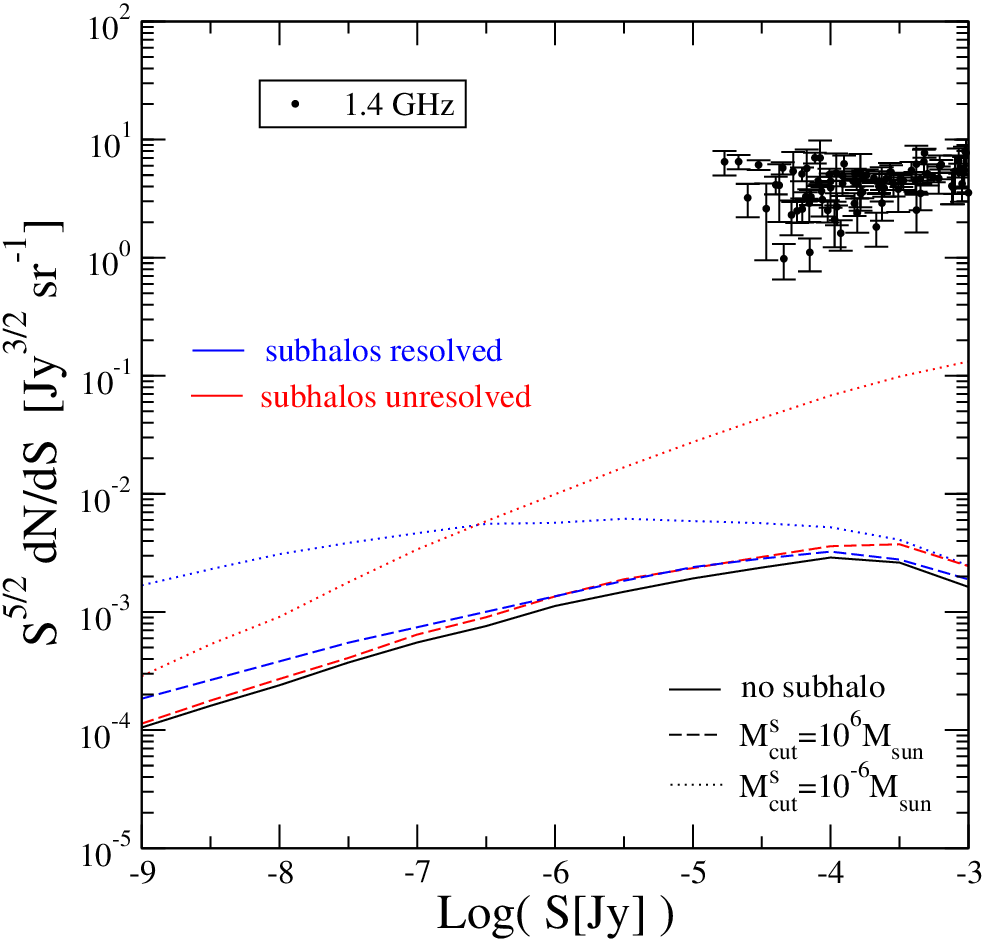}
 \end{minipage}
 \ \hspace{1mm} \
 \begin{minipage}[htb]{0.3\textwidth}
   \centering
   \includegraphics[width=\textwidth]{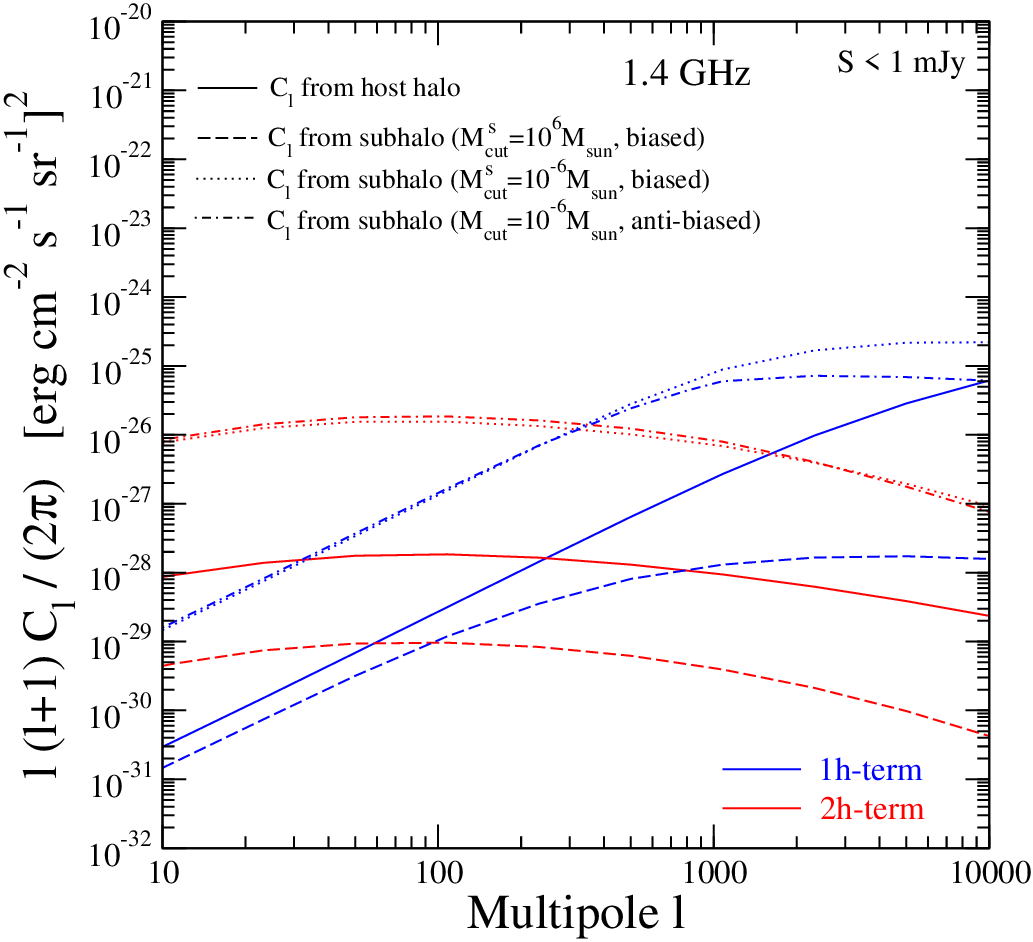}
 \end{minipage}
    \caption{{\bf Substructures.} {\it Left}: Total intensity for the benchmark annihilating model B1 with (dashed/dotted) and without (solid) substructure contribution. The substructure model is such that 10\% of total mass is in substructures with subhalo mass function $dn_s/dM_s\propto M_s^{-1.9}$. The minimum subhalo mass is $M_{cut}^s= 10^6 M_{\odot}$ (dashed) or $M_{cut}^s= 10^{-6} M_{\odot}$ (dotted). {\it Central}: Number counts at 1.4 GHz for the same models of left panel. Blue color stands for the case in which substructures are considered as single sources, while red denotes the picture in which their contribution is only a boost of the signal from the parent halo. {\it Right:} Angular power spectra of host-halos and sub-halos for the same two models of left panel. In the model with $M_{cut}^s= 10^{-6} M_{\odot}$, both biased and antibiased radial distributions of subhalos are shown. }
\label{fig:sub}
 \end{figure}

To understand the possible contribution from substructures, we consider a model such that 10\% of total mass is in substructures with subhalo mass function $dn_s/dM_s\propto M_s^{-\alpha}$ and $\alpha=1.9$ which is in agreement with simulations for Milky Way size halos~\cite{Madau:2008fr,Springel:2008cc} (see also similar treatments in, e.g., Ref.~\cite{Ando:2006cr,Fornasa:2009qh}), while all other clustering details are kept as in the benchmark case of the previous Section. 
The minimum subhalo mass is taken to be either $M_{cut}^s= 10^6 M_{\odot}$ or $M_{cut}^s= 10^{-6} M_{\odot}$ (with the minimum mass for the parent halo fixed to $M_{cut}=10^6 M_{\odot}$).
Those two models leads to a boost in the total intensity of a factor of 1.4 and 7, respectively, see Fig.~\ref{fig:sub}a.

For what concerns number counts, we have to distinguish the case in which a subhalo is identified as a single source to the case such that the subhalo only contributes to boost the signal of the parent halo. In the first case, we say the subhalo is resolved, while in the second it is unresolved.
To bracket uncertainties related to the possibility of resolving substructures, we consider two extreme cases such that subhalos are either all resolved or all unresolved. 
The first possibility leads to a flattening in the number count spectrum, since the number of sources at lower brightness is increased.
On the contrary, the second possibility steepen the spectrum, since the boost from subhalos pushes parent halos to larger brightnesses.

Fig.~\ref{fig:sub}c shows the angular power spectra of the host--halos and subhalos in the scenario where both contributions are included\footnote{For simplicity, we only consider the autocorrelation terms of host--halo and subhalo, disregarding the cross-correlation power spectrum (which is important only if the two contributions are of the same order).}. The host--halo case is simply rescaled by a factor $(1-f)^4$ (with $f=0.1$) with respect to the plot of Fig.~\ref{fig:DMdistr}c. 
With the formalism provided by Eqs.~(\ref{eq:cl}), (\ref{eq:cl1h}), (\ref{eq:cl2h}) and (\ref{eq:lumDMa}), the subhalo contribution is computed in a similar fashion as for the main halo case, but with each main structure now ``weighted'' by different luminosity and spatial profile that follow from the (averaged) substructure distribution. 
The emission profile is now shallower than $\rho^2$, and the emission from largest halos is enhanced since they host largest number of subhalos. 
The latter property implies that the relative significance of the one--halo term with respect to the two--halo term increases if smallest substructures are not introduced (case with $M_{cut}^s= 10^6 M_{\odot}$). 
The first property leads instead to a flattening of the power at very large multipoles in the one--halo term with respect to the host--halo case.
The effect is visible for subhalo density going as $\propto\rho$ (biased case), and even more pronounced in the antibiased case~\cite{Madau:2008fr}, where the radial distribution of subhalos is much flatter than that of the dark matter in the host halo.

\subsubsection{Dark matter microscopic properties}

\begin{figure}[t]
 \begin{minipage}[htb]{0.3\textwidth}
   \centering
   \includegraphics[width=\textwidth]{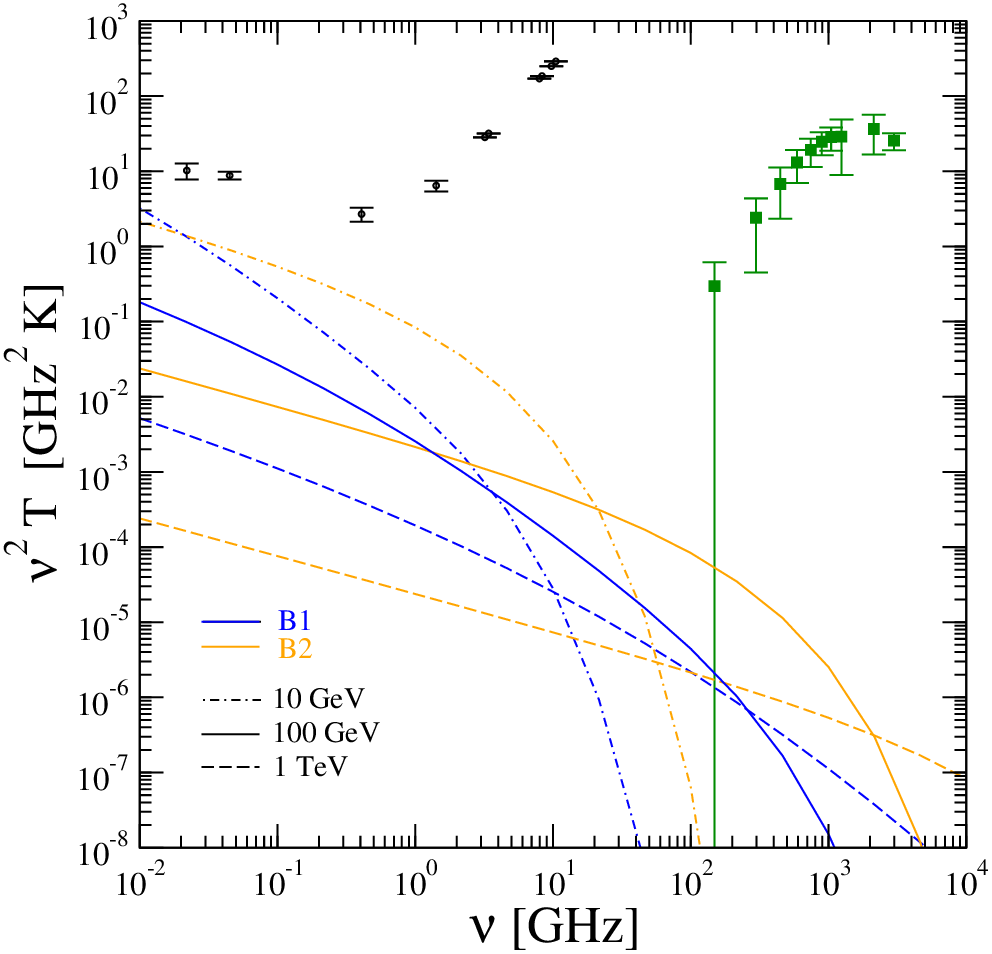}
 \end{minipage}
 \ \hspace{2cm} \
 \begin{minipage}[htb]{0.3\textwidth}
   \centering
   \includegraphics[width=\textwidth]{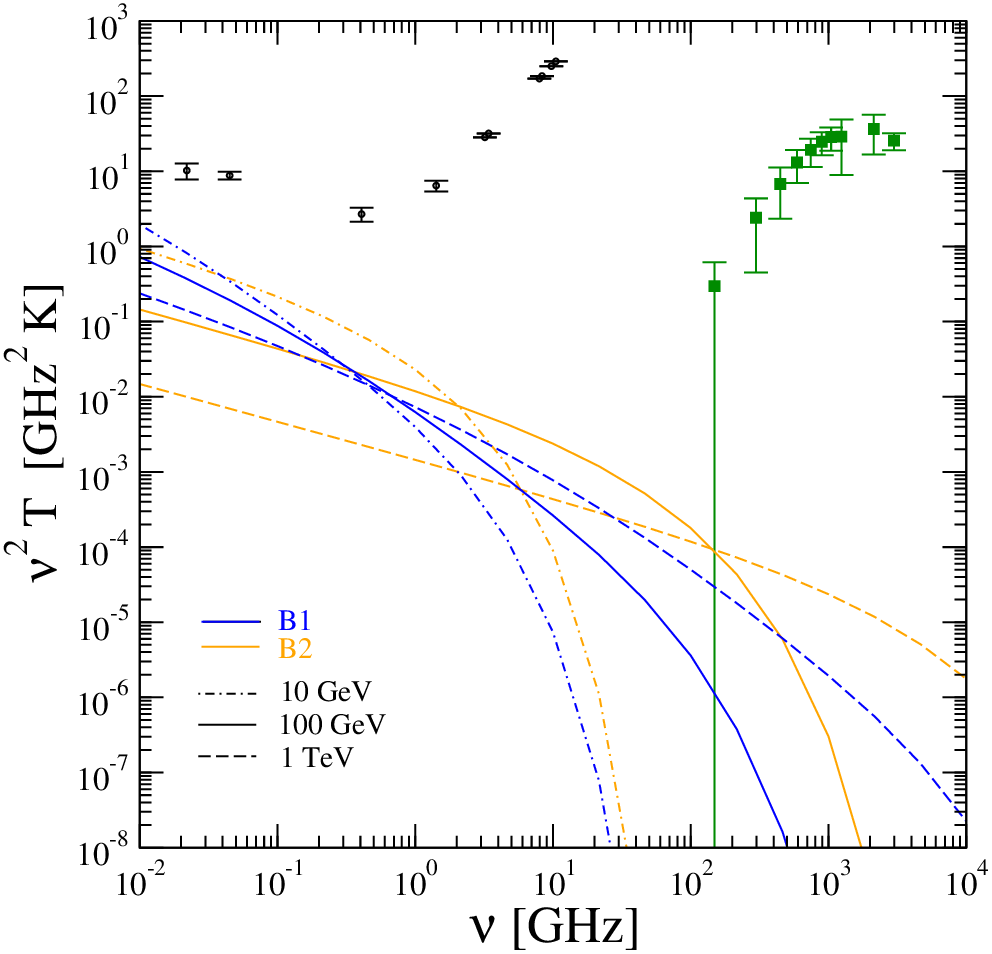}
 \end{minipage}
    \caption{{\bf Microscopic properties.} Annihilating cases with $\sigma_a v=3\cdot10^{-26}\, {\rm cm^3s^{-1}}$ (left), and decaying cases with $\tau=10^{28}$ s (right).}
\label{fig:microprop}
 \end{figure}

In Fig.~\ref{fig:microprop}, we show how the synchrotron signal depends on the DM microscopic properties.
We consider few different masses, and two rather different final states of annihilations/decays, $\bar b-b$ and $\mu^+-\mu^-$, the first inducing a $e^-/e^+$ spectrum much softer than the latter (and in turn a synchrotron spectrum with analogous properties, see the plot).

The dependence on annihilation/decaying rate is straightforward since it is just a normalization parameter\footnote{It could actually affect small--scale clustering, but only for very dense regions and large rate which are not considered here.}.

In the monochromatic approximation of synchrotron radiation, the energy of an electron emitting at frequency $\nu$ is given by $E\simeq 15\sqrt{\nu_{GHz}/B_{\mu G}}$ GeV (where $\nu_{GHz}$ is the frequency in GHz and $B_{\mu G}$ is the magnetic field in $\mu G$). Assuming a magnetic field in the reasonable range of few microGauss, we can get an estimate of the energy of electrons/positrons involved in the emission at a certain frequency.
With this simple relation we can understand how to relate the $e^+-e^-$ spectrum of annihilation/decay with the emission. 
The threshold in Fig.~\ref{fig:microprop} is given by the fact that no electrons can be produced with energy above the mass $M_{\chi}$ (in the annihilating case) or above $M_{\chi}/2$ (in the decaying case), since WIMPs are non--relativistic. Then, softer is the $e^+-e^-$ spectrum of annihilation/decay, smaller is the energy at which the emission starts dropping.

The normalization of the annihilation signal scales as the square of the number density, so as $M_{\chi}^{-2}$, while in the decaying case as $M_{\chi}^{-1}$.
Therefore the spread in the decaying case is much smaller, and it is more predictive than the annihilating case.

For the same mass and annihilation/decaying rate, WIMPs annihilating into $\mu^+-\mu^-$ induce a signal which is brighter (fainter) at high (low) frequencies with respect to the $\bar b-b$ case.  
We have already stressed that the bulk of the DM induced emission is generated at very low redshift. Therefore there is no particular mixing among different redshift slices and the impact of features of the annihilation/decay spectrum on source counts and anisotropies is similar to what shown for the total intensity.

The comparison between annihilating and decaying DM has been performed in Figs.~\ref{fig:counts}b, \ref{fig:cl}b, and \ref{fig:cl}c. The $\rho^2$ scaling in the former case enhances the power on small scales and the number of faint sources with respect to the latter case, which scales with $\rho$.
These facts makes the emission in the annihilating case more distinguishable from other astrophysical radio sources.

\subsubsection{Magnetic field}
\label{sec:B}
\begin{figure}[t]
 \begin{minipage}[htb]{0.3\textwidth}
   \centering
   \includegraphics[width=\textwidth]{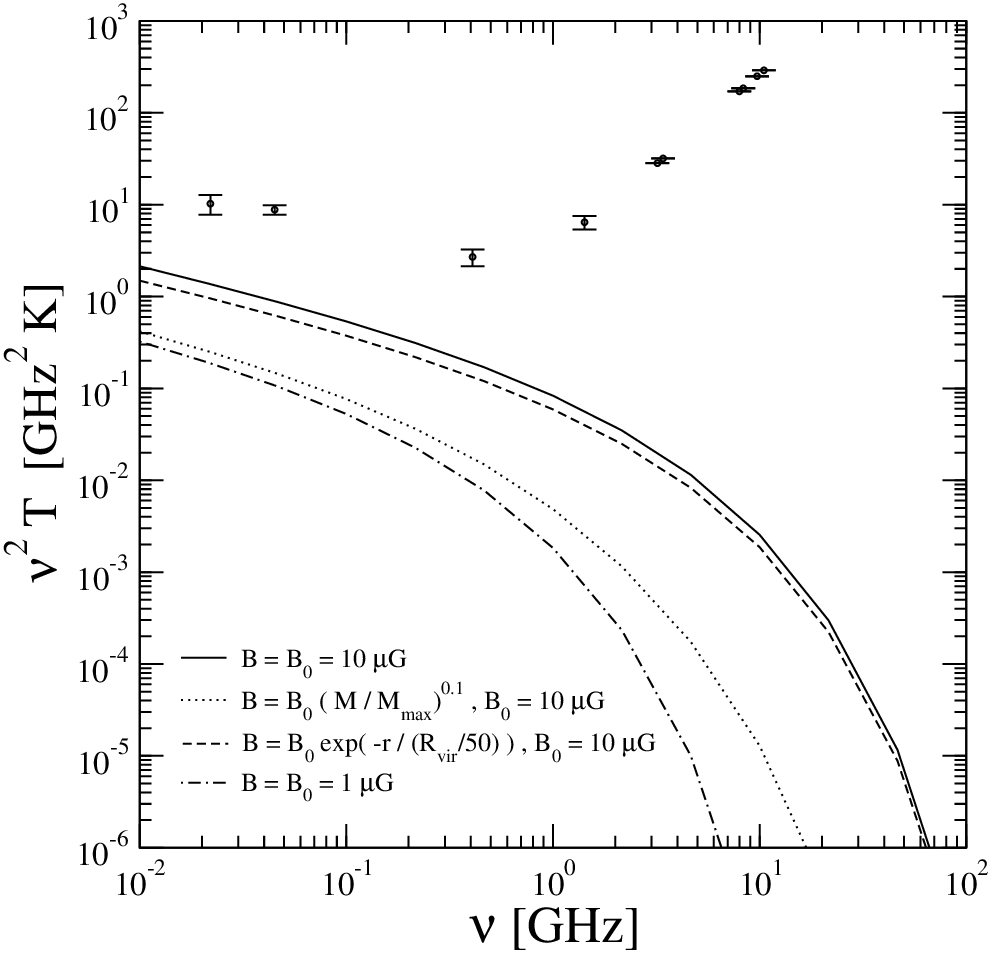}
 \end{minipage}
 \ \hspace{1mm} \
 \begin{minipage}[htb]{0.3\textwidth}
   \centering
   \includegraphics[width=\textwidth]{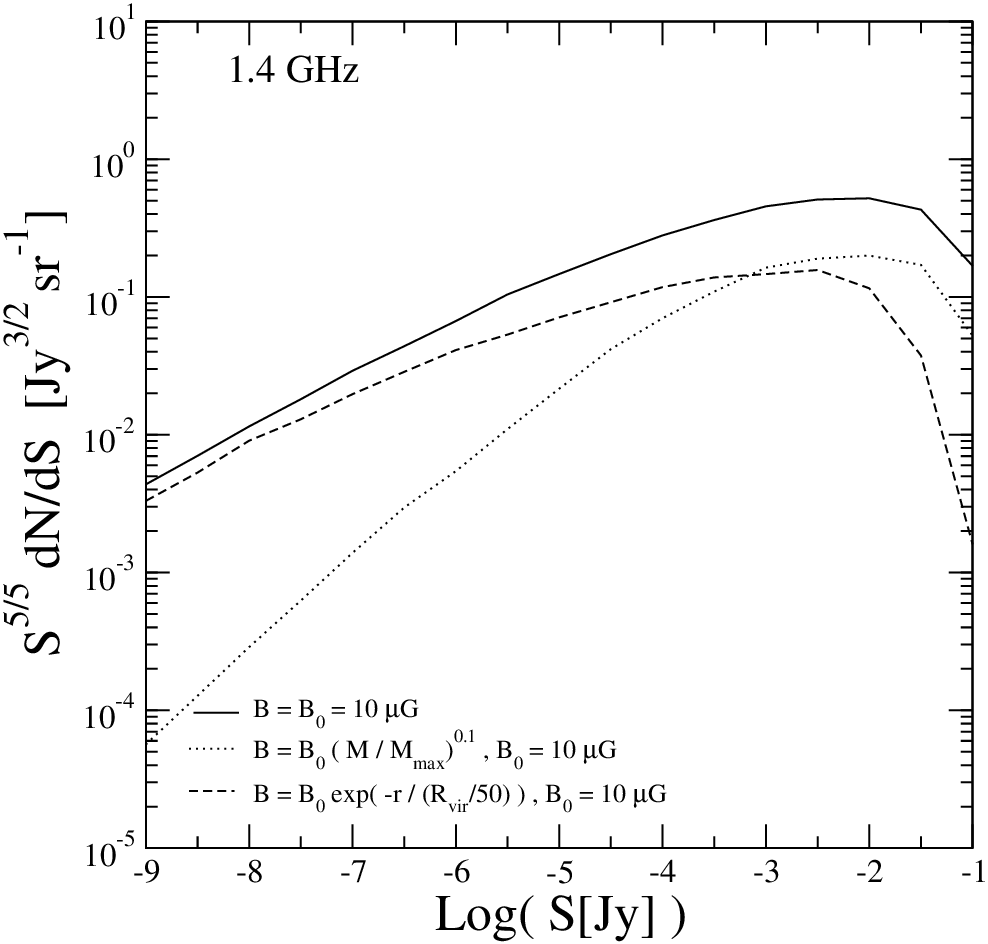}
 \end{minipage}
 \ \hspace{1mm} \
 \begin{minipage}[htb]{0.3\textwidth}
   \centering
   \includegraphics[width=\textwidth]{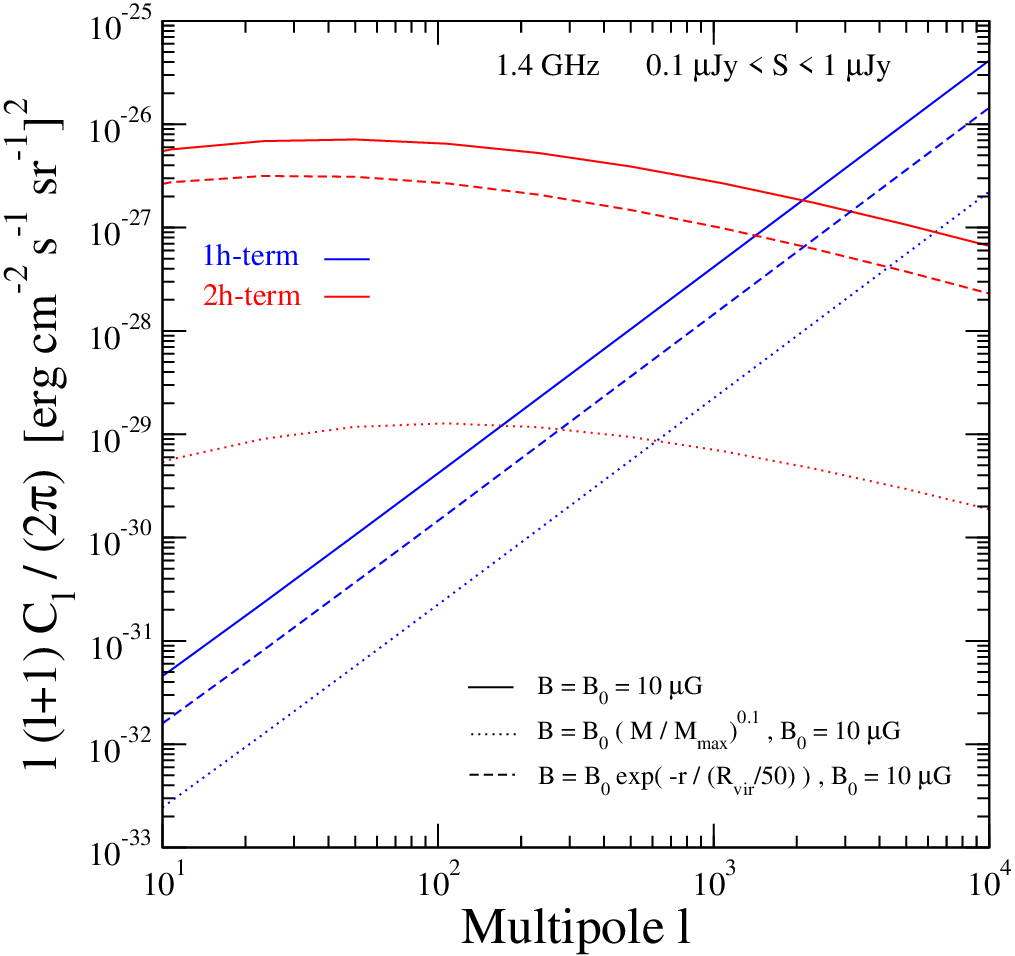}
 \end{minipage}
    \caption{{\bf Magnetic field.} {\it Left}: Total intensity induced by the benchmark annihilating model B2 considering four different assumptions for the magnetic field strength. {\it Central}: Number counts at 1.4 GHz for the same models of left panel.  {\it Right:} Angular power spectra of DM sources with brightness $0.1\,\mu {\rm Jy}<S<1\, \mu {\rm Jy}$ at 1.4 GHz for the same models of left and central panels. One-halo and two-halo terms of $C_\ell$ are shown in blue and red, respectively. }
\label{fig:B}
 \end{figure}

First of all, we investigate the impact of the strength of the magnetic field, but still assuming $B$ to be constant in time and space.
The magnetic field affects both energy losses and synchrotron power. 
As mentioned above, we consider energy losses due to synchrotron and inverse Compton scattering on CMB. 
The case of IC scattering on CMB is equivalent to an energy loss due to synchrotron radiation on a magnetic field of $B_0\simeq3\, \mu$G (for low energy electrons, i.e., neglecting Klein--Nishina corrections). 
Therefore for magnetic fields much larger than $B_0$, the dominant energy loss is synchrotron.
In this case, the total flux is only marginally affected by the magnetic field strength. 
Indeed a decrease in $B$ leads to a lower synchrotron power which is however compensated by the larger cooling time, and vice--versa.
This is true provided that the confinement time is large and that the energy of electrons needed for the emission at a certain frequency is far from the threshold of the spectrum, which is (a fraction of) the WIMP mass. 
We have already mentioned that, for a magnetic field of few $\mu$G, emissions at radio frequencies are mostly generated by electrons with energy around 1--10 GeV.
For magnetic fields smaller than $3\, \mu$G, the scaling is instead roughly proportional to $B^2$, since only the synchrotron power is affected, while energy losses do not significantly vary. The case with $B=1\,\mu$G is plotted in Fig.~\ref{fig:B}a which shows that, with respect to the case with $B=10\mu$G, there is a factor of 10 in the normalization, and the threshold occurs for lower frequencies.

To understand what could be the effects of a more realistic treatment of magnetic field, we consider a simple but instructive analytical form for $B$:
\be
B(M,r,z)=B_0\,\left(\frac{M}{M_{\rm max}}\right)^a\,(1+z)^{-b}\, \exp{\left(-\frac{r}{c\,R_v}\right)}\;,
\label{eq:B}
\ee
with $M_{\rm max}=10^{18}M_{\odot}$ (the exact chosen value doesn't really matter) being the largest halo mass considered.
With this simple expression, we keep spherical symmetry and the crucial parameter to set the strength for a given source is again the mass $M$, so Eqs.~\ref{eq:intgen}-\ref{eq:PIC} in Section II are still valid and easily solvable (and will be exploited in the rest of the paper, no longer resorting to the simplified case of Eqs.~\ref{eq:Iannsimp}-\ref{eq:Cldecsimp}). On the other hand, Eq.~(\ref{eq:B}) allows us to study a magnetic field which increases with the size of the source ($a\geq 0$ following observations of larger $B$ in clusters with respect to galaxies), which has a time evolution (typically $b\geq 0$), and which decreases in the outer part of the source far from the luminous region ($c\leq 1$). 

The case considered so far is simply recovered by $(a,b,c)=(0,0,\infty)$.
As shown in Figs.~\ref{fig:totI}c and \ref{fig:counts}c, the bulk of the DM signal is emitted at very low redshift. Therefore $b\neq0$ leads to negligible differences, unless considering extreme evolution: the $z$--dependence of the magnetic field can therefore be safely neglected.
The scalings with object mass and spatial extension of $B$ in the object can be instead more interesting and in Fig.~\ref{fig:B} we show few examples, considering $a=0.1$ (to go from $B\sim10\,\mu$G in large systems to $B\sim1\,\mu$G in small galaxies) and $c=1/50$ (e.g., in our Galaxy the virial radius is $R_v\sim250$ kpc and the magnetic region is few kpc in the vertical direction).
The case with $a>0$ leads to a suppression in the emission from small objects, which implies a depletion in the number counts at small brightness (Fig.~\ref{fig:B}b) and an overall depletion in the total intensity (Fig.~\ref{fig:B}a). The one--halo term of angular power spectrum increases its relative importance with respect to the two--halo term, and the crossing point between the two curves in Fig.~\ref{fig:B}c occurs at smaller $\ell$ compared to the $a=0$ case.
The model with $c\neq0$, leads instead to a suppression in the emission from large objects. Indeed, since the concentration parameter $c_{vir}$ decreases as the halo mass increases and thus the DM halo profile drops more quickly in smaller objects, the impact of cutting the emission at a certain radius $\ll R_v$ is larger on massive systems. The effect on total intensity and angular correlation is however modest, with a very mild increase of the relative importance for the two--halo term.

\subsubsection{Electron/positron transport in extra--galactic sources}
\label{sec:diff}

So far we have been considering a picture such that the $e^-/e^+$ produced in WIMP annihilations/decays radiate at the same time and place of injection, and that all the power injected by DM is radiated. In other words, we have been using the solution to the transport equation for $e^-/e^+$ given by Eq.~\ref{eq:transp1}.
The time-scale associated to energy losses is taken to be:
\be
 \tau_l\simeq\frac{3.7\cdot 10^{16}\,{\rm s}}{(1+z)^4+0.095\,B_{\mu G}^2}\frac{{\rm GeV}}{E}\;.
\label{eq:transp2}
\ee
We included energy losses from synchrotron and inverse Compton on CMB (here reported neglecting Klein--Nishina corrections for IC, which is a good approximation up to 10 TeV for scattering with CMB photons).
Since we considered magnetic fields of 1--10 $\mu$G and a radio emission which is associated to electrons with energy 1--10 GeV, we typically have $\tau_l\lesssim10^{16}$ s.
Three are the conditions that need to be satisfied such that Eq.~(\ref{eq:transp1},\ref{eq:transp2}) provides a good approximation of the real picture, namely $\tau_l$ has to be, in each structure, significantly smaller than timescales associated to: 1) evolution of the structure, 2) diffusion of charged particles and 3) confinement of charged particles.

Condition 1) is the one that allows us to adopt a time--independent solution.
Since $\tau_l$ is always much smaller than the Hubble time (as already noticed in Ref.~\cite{Profumo:2009uf}), we can assume that $e^-/e^+$ are injected at the same redshift at which they emit and neglect (in this respect) the cosmological evolution. Also, since the emission is concentrated at very low redshift (and span a very limited redshift range), we can assume $e^-/e^+$ to have reached equilibrium and consider a steady--state solution. 

Concerning condition 2), we first notice that so far the spatial distribution of $e^-/e^+$ in the source is assumed to follow from the DM spatial profile. This can be considered as a good approximation only if the diffusion scale $d_L\simeq\sqrt{4\,D\,\tau_l}$ (where $D$ is the diffusion coefficient $D=D_0\,(E/{\rm GeV})^\delta$) is smaller than (comparable to) the scale at which the DM profile shows significant variations.
If we consider the value of $D_0=3\cdot 10^{28} {\rm cm^2/s}$ as inferred in the Milky Way, we have $d_L$ larger than few kpc. Therefore, in particular for cuspy profiles, the spatial diffusion has an impact on the computed signal. It tends to smooth out the cusp of the injection distribution into shallower profiles.
However, provided that the particles do not efficiently escape the object (see discussion below) and the magnetic field does not dramatically vary within the diffusive region, the brightness of single sources is not affected (and in turn also the total extragalactic intensity). Only the angular correlation can be modified at very small scales, in a way that can be understood moving from the Moore or NFW case to the Burkert curve in Fig.~\ref{fig:DMdistr}c. Indeed, effectively, a cored profile mimics the spatial distribution of $e^-/e^+$ injected with cuspy profile and then undergoing spatial diffusion. 

\begin{figure}[t]
 \begin{minipage}[htb]{0.3\textwidth}
   \centering
   \includegraphics[width=\textwidth]{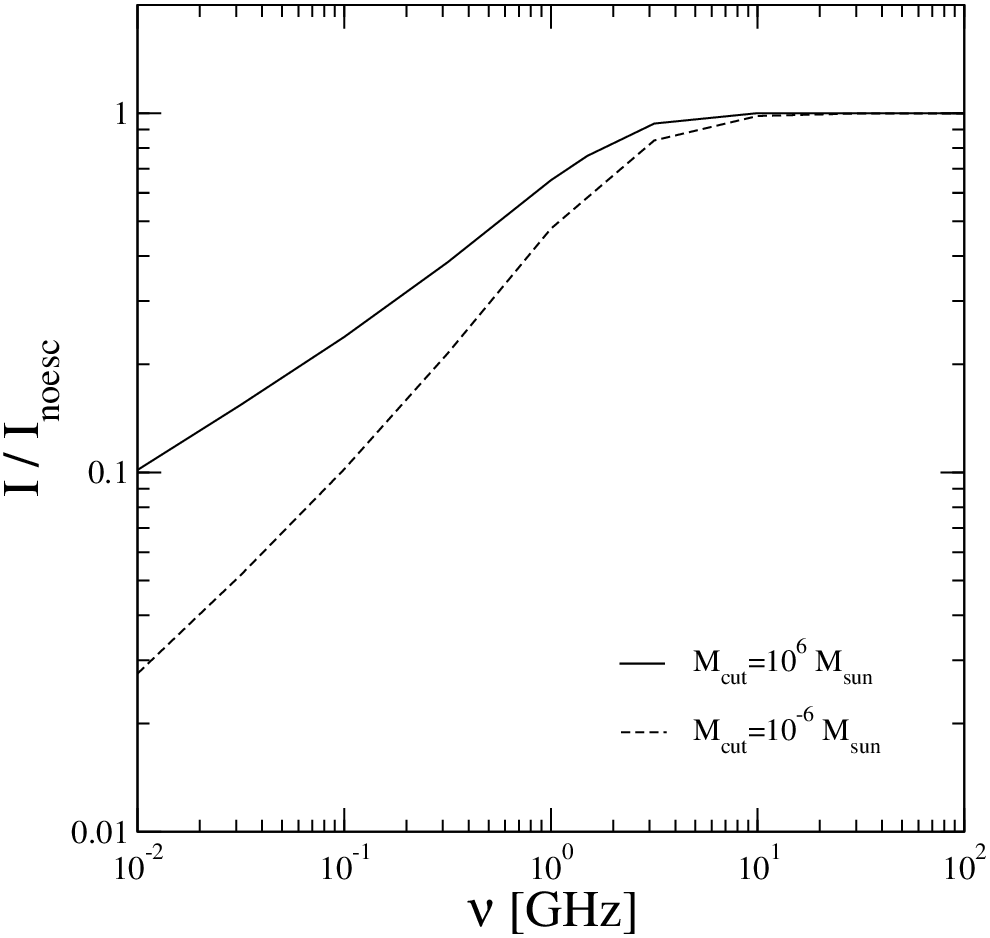}
 \end{minipage}
 \ \hspace{1mm} \
 \begin{minipage}[htb]{0.3\textwidth}
   \centering
   \includegraphics[width=\textwidth]{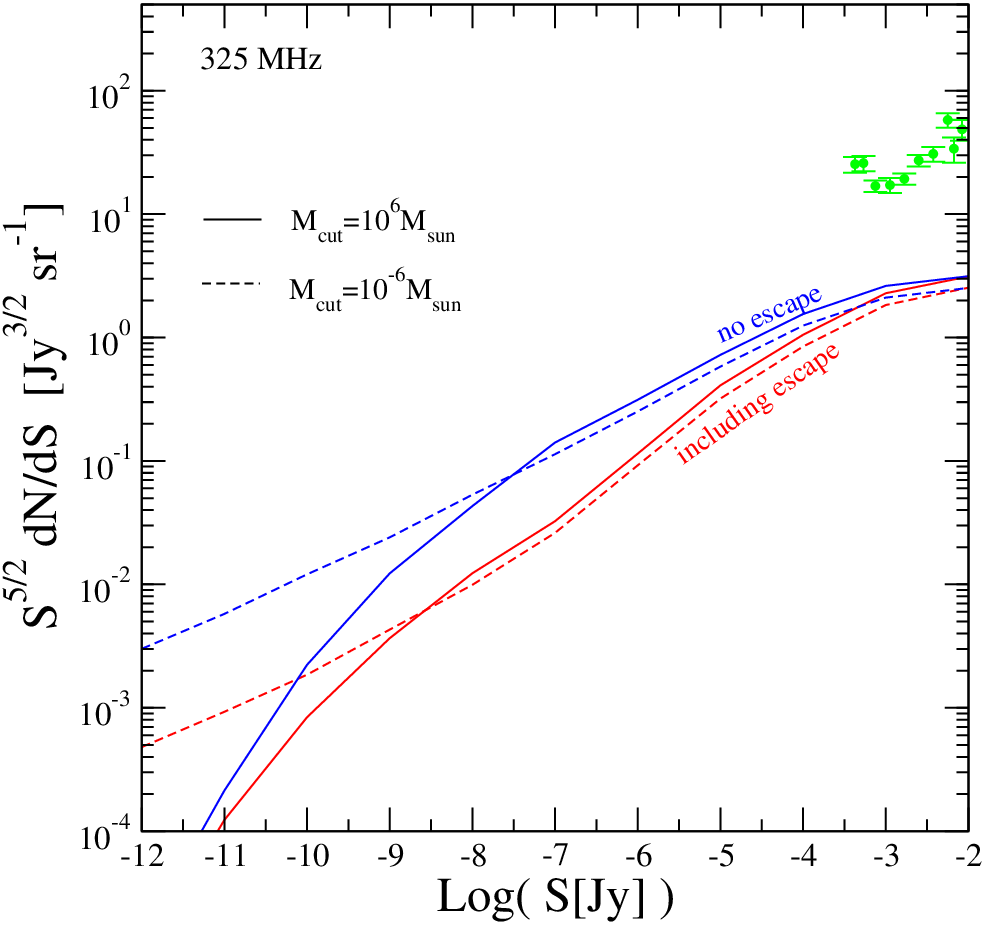}
 \end{minipage}
 \ \hspace{1mm} \
 \begin{minipage}[htb]{0.3\textwidth}
   \centering
   \includegraphics[width=\textwidth]{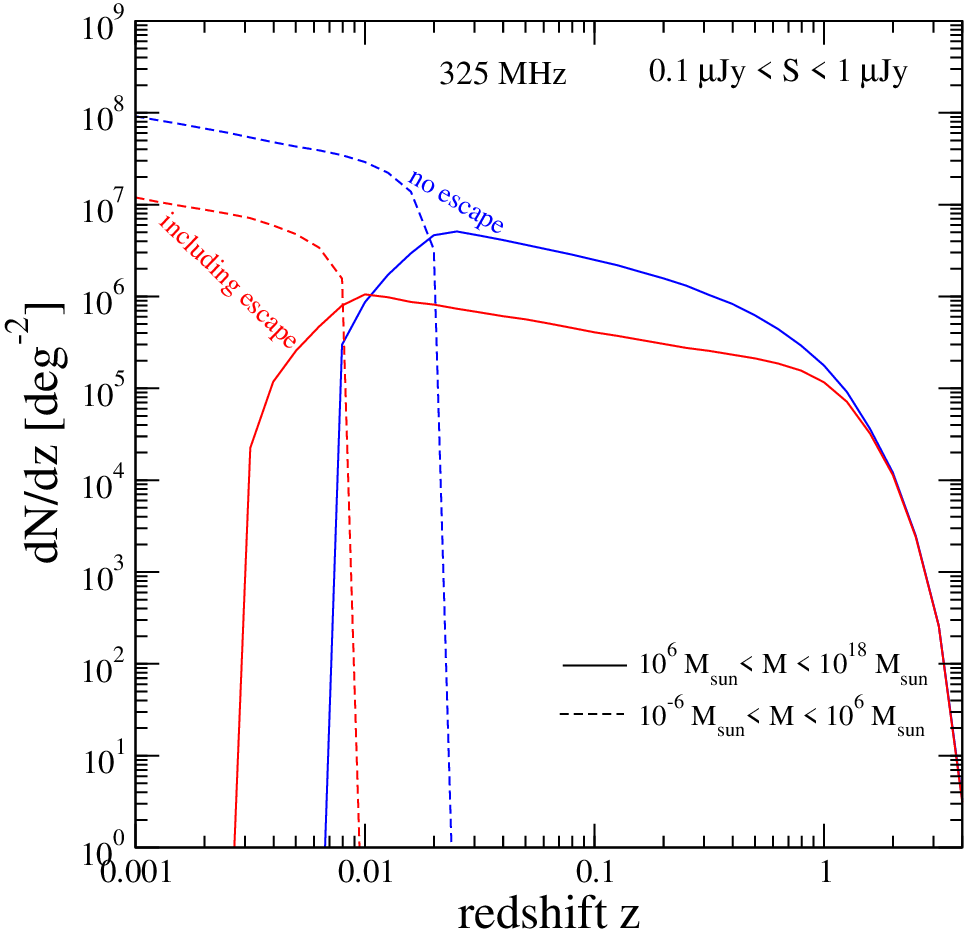}
 \end{minipage}
    \caption{{\bf Confinement.} {\it Left}: Comparison of total intensity with/without possible escape of $e^-/e^+$ after injection from DM annihilation (in the benchmark annihilating model B2). The two cases with $M_c=10^{6} M_{\odot}$ and $M_c=10^{-6} M_{\odot}$ show that small structures and low frequencies can be significantly affected. {\it Central}: Source counts at 325 MHz with/without possible escape of $e^-/e^+$ and considering $M_c=10^{6} M_{\odot}$ and $M_c=10^{-6} M_{\odot}$ (again in the benchmark annihilating model B2).  {\it Right:} Redshift distribution of number counts for faint DM sources $0.1\,\mu {\rm Jy}<S<1\, \mu {\rm Jy}$. Small ($10^{-6} M_{\odot}<M<10^{6} M_{\odot}$) and large ($10^{6} M_{\odot}<M<10^{18} M_{\odot}$) halos are shown separately.}
\label{fig:escape}
 \end{figure}

If condition 3) is not verified the source brightness can be depleted. Indeed it means that $e^-/e^+$ can escape the object before radiating at the frequency of interest.
A full description of spatial diffusion in all the extragalactic sources is obviously beyond the goal of this paper.
To estimate the impact of a finite escape length, we multiply the $e^-/e^+$ spectrum $dN_e/dE$ by a factor $\exp(-\tau_{cool}/\tau_{conf})$.
The confinement time $\tau_{conf}$ can be estimated in a simple leaky box model as $\tau_{conf}(E)\simeq L^2/(2\,D(E))$, where $L$ is the size of the diffusive region.
The cooling time is instead $\tau_{cool}(E)\simeq \tau_l(E)(E/E_p-1)$, where $E_p$ is the electron energy corresponding to the peak in the synchrotron emission at the frequency of interest.
Under such approximations, the confinement time in the Milky Way is $\tau_{conf}\simeq10^{15}(E/{\rm GeV})^{-\delta}$ s (having taken $L$ to be few kpc).
Therefore even a relatively large structure as our Galaxy can suffer of a depletion in the emission at low frequency (since the latter involves electrons with lower energies which means larger cooling times and diffusion lengths) due to the escape of a fraction of the injected $e^-/e^+$.
Note, however, that we have been neglecting the energy loss due to IC on starlight which, in high-luminosity structures, can become relevant and reduce the cooling time.  
The effect of electron escape becomes more and more important for smaller objects, and the possible large enhancement provided by DM halos with masses down to $10^{-6} M_{\odot}$ (see Fig.~\ref{fig:DMdistr}a) can be partially erased by the efficient escape.
This is shown in Fig.~\ref{fig:escape}a, where we compare the total intensity of the case neglecting electron escape to the case where it is accounted for in the described simplified picture (for all the plots in Fig.~\ref{fig:escape}, we choose $L=R_v/50$ and $D=3\cdot 10^{28} (E/{\rm GeV})^{1/3} {\rm cm^2/s}$). 
At frequency above 10 GHz, the effect is negligible since the emission mainly comes from electrons with energy larger than 10 GeV which loose energy very effectively and radiate nearly instantaneously after injection.
Fig.~\ref{fig:escape}b shows that number counts associated to small objects at low brightness are depleted (up to extremely low brightness where the counts stem from sources at high--redshift having much shorter diffusion lengths due to the much larger IC cooling). 
In Fig.~\ref{fig:escape} we plot how the redshift distributions of large ($10^{6} M_{\odot}<M<10^{18} M_{\odot}$) and small ($10^{-6} M_{\odot}<M<10^{6} M_{\odot}$) sources are affected. In the latter case, the electron escape largely depletes the emission and also narrows down the contribution to lower redshift.
Larger objects are too bright to emit at brightness $0.1\,\mu {\rm Jy}<S<1\, \mu {\rm Jy}$ for redshift $z<0.01$ if the escape is neglected. When the latter is included, however, the contribution from sources in the lighter end of the mass range is shifted towards lower redshift. Most massive halos are dominant at large redshift and their emission is unaffected.

The effect of electron escape on angular correlation at low brightness is to mildly enhance the relative power of large versus small scales.
This is due to the depletion of the contribution from small sources and to the shift towards lower redshift of intermediate--mass sources.

\subsection{Constraints on WIMP parameter space}
\label{sec:constraints}
Having discussed the impact of the main uncertainties on the WIMP induced signals, we can now proceed to derive constraints on the WIMP parameter space. 
They are obtained through the comparison with the data presented in Section 2. We conservatively choose not to add any other astrophysical contribution to the theoretical flux.

\begin{figure}[t]
 \begin{minipage}[htb]{0.3\textwidth}
   \centering
   \includegraphics[width=\textwidth]{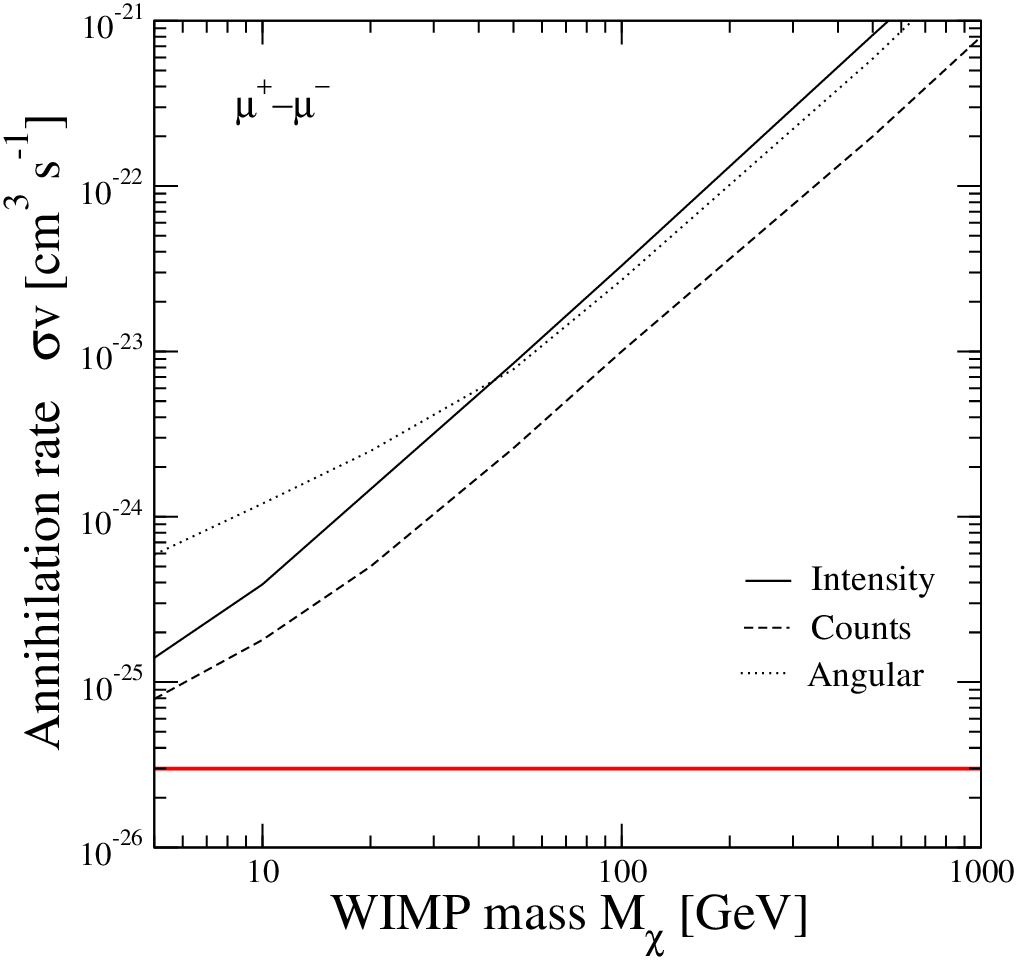}
 \end{minipage}
 \ \hspace{1mm} \
 \begin{minipage}[htb]{0.3\textwidth}
   \centering
   \includegraphics[width=\textwidth]{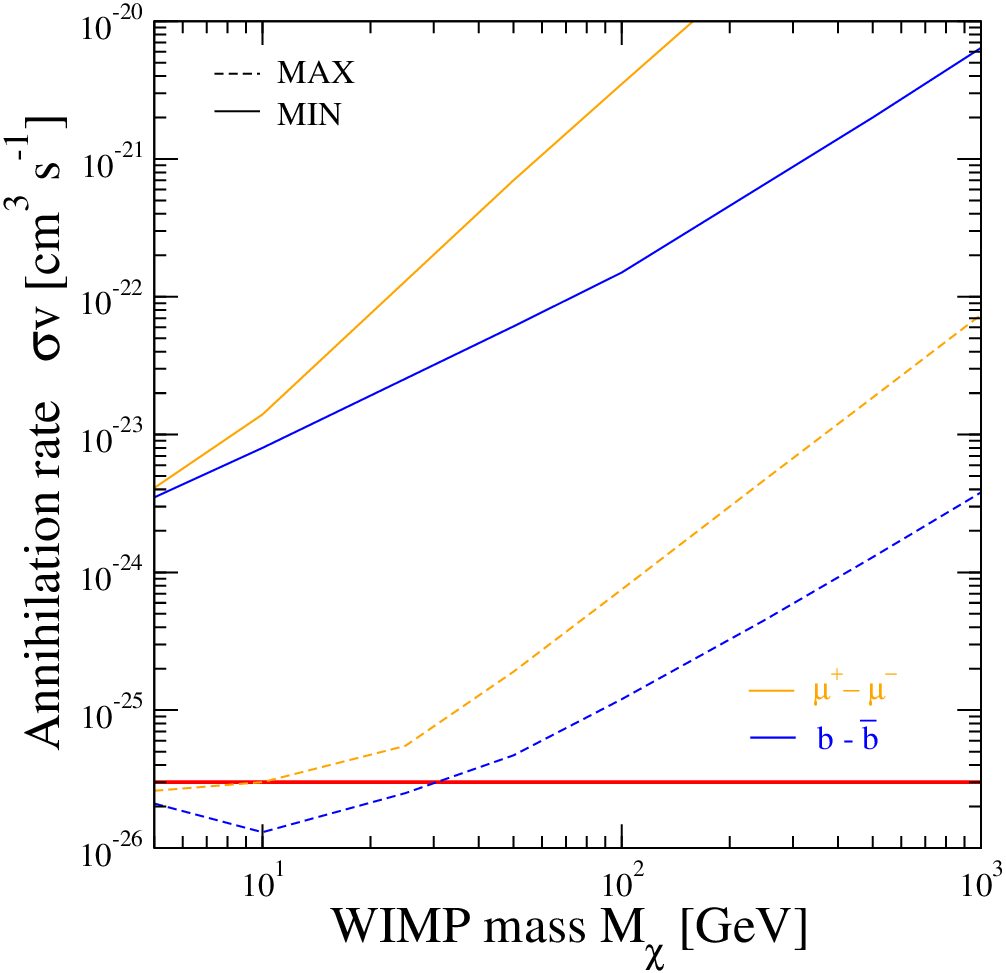}
 \end{minipage}
 \ \hspace{1mm} \
 \begin{minipage}[htb]{0.3\textwidth}
   \centering
   \includegraphics[width=\textwidth]{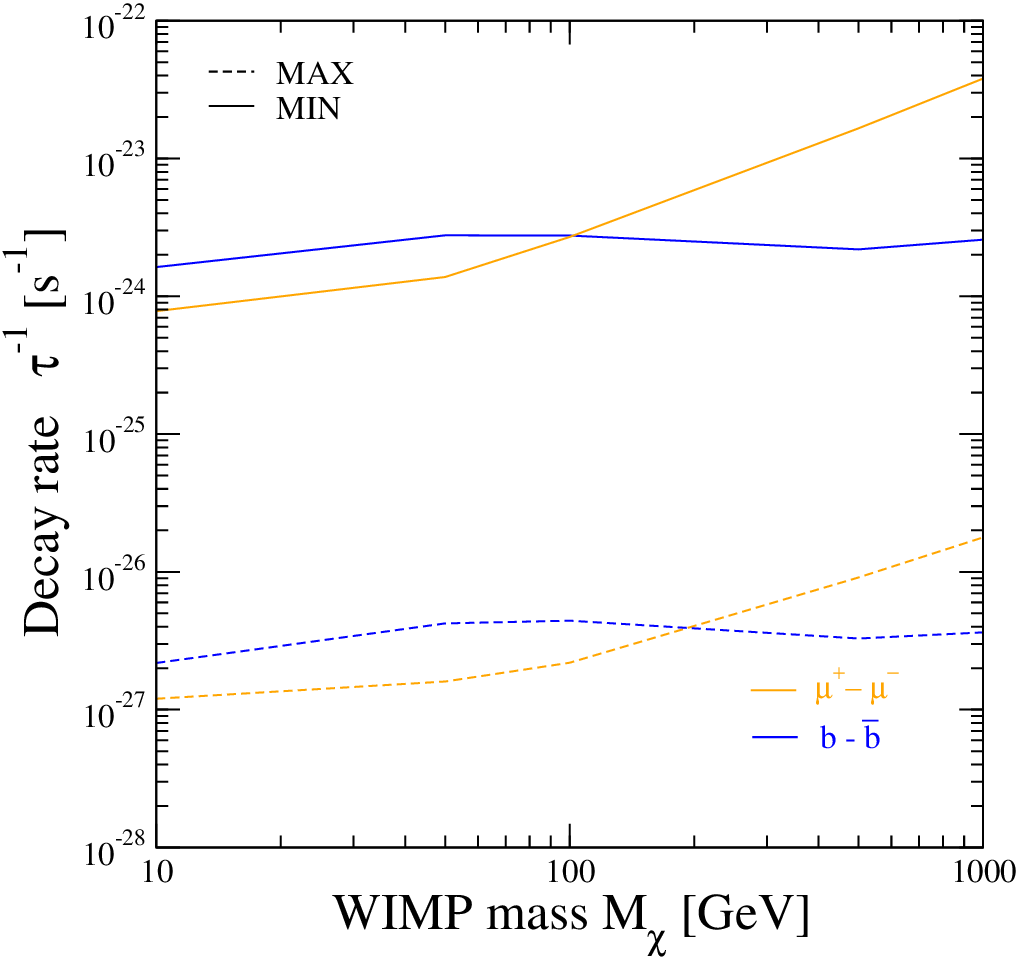}
 \end{minipage}
    \caption{{\bf Constraints.} {\it Left}: Bounds from total intensity (solid), source counts (dashed), and angular correlation (dotted) for WIMPs with dominant annihilation channel $\mu^+-\mu^-$. Clustering and other astrophysical assumptions are as described in Sec.~\ref{sec:bench}. The reference `thermal' annihilation rate $\sigma_a v=3\cdot10^{-26}\, {\rm cm^3s^{-1}}$ is shown for comparison. {\it Central}: Constraints on the parameter space of WIMP annihilating models in the MIN (solid) and MAX (dashed) scenarios (see text for details). Final states of annihilations are $b-\bar b$ (blue) and $\mu^+-\mu^-$ (orange). {\it Right:} Same of central panel, but for decaying DM.  }
\label{fig:mvssv}
 \end{figure}

Considering the benchmark scenario depicted in Sec.~\ref{sec:bench}, we compare bounds from total intensity, source counts, and angular correlation in Fig.~\ref{fig:mvssv}a. 
The exclusion curves scales roughly as $M_\chi^2$ for $M_\chi$ above few tens of GeV, while for lighter WIMPs the induced spectra can become significantly softer than data, which leads to a flattening of the curves. 
Note that the total--intensity bound is subdominant. However, as already mentioned, we have been exploiting extragalactic estimates from the ARCADE collaboration which are 5--6 times larger than the expected emission. Whether this excess will be ascribed to instrumental or galactic contaminations rather than to extragalactic sources, the bound should be accordingly rescaled.
Moreover, for different scenarios with an enhanced total emission due to the inclusion of low brightness objects (e.g., reducing $M_{cut}$ or introducing (resolved) substructures), the intensity constraint becomes more effective than number counts since for the latter there are no data at low brightness.

As already stressed above, the experimental data currently available make constraints from anisotropies and number counts significantly weaker with respect to what one could in principle achieve with deeper future survey (while the total--intensity bound is not expected to dramatically improve).

On the other hand, and again as discussed above, since DM halos can host much brighter baryonic components, the induced emission of the former can be observationally hidden by the latter for what concerns measurements of number counts and two--point angular correlation function. 
We leave the discussion on how to separate the DM contribution for a further work.
Here, to derive conservative constraints we consider only total intensity and angular power spectra $C_\ell$ derived from observations with single--dish telescopes and CMB satellites (which get contributions from sources at all brightnesses). The lowest multipole considered at low (high) frequency is $\ell=50$ ($\ell=3000$) because of contaminations from galactic (CMB) anisotropies which make any estimates of extragalactic source contribution on larger scales too uncertain~\cite{Giardino,LaPorta:2008ag,Regis:2011ji}.

For our final constraints on WIMP parameter space, we define two sample cases bracketing the effect of quantities which affect the extragalactic DM radio flux but {\it not} directly related to DM microscopic properties (namely: clustering, substructures, magnetic field, and spatial diffusion). The first scenario (called MIN) is such that the flux is minimized, while the second (MAX) corresponds to the most optimistic assumptions. 
In both models, we adopt the halo mass function from Ref.~\cite{Sheth:1999mn}, the concentration of halos from Ref.~\cite{MunozCuartas:2010ig}, and the DM distribution in halos follows a NFW profile~\cite{Navarro:1996gj}. 

In the MIN case, we set $M_{cut}= 10^6 M_{\odot}$, no contributions from substructures, magnetic field as in Eq.~(\ref{eq:B}) (with $B_0=10\,\mu$G and $(a,b,c)=(0.1,0,1/50)$, and electron escape as described in Sec.~\ref{sec:diff}.
For the MAX case, we choose $M_{cut}= 10^{-6} M_{\odot}$, $B=10\,\mu G$, and $e^+/e^-$ are assumed to radiate at the same place where they are injected (while the inclusion of substructures makes little difference unless considering extreme models for the subhalo mass function).
 
We choose again the two benchmark final states of annihilations (decays), $b-\bar b$ (inducing a softer $e^+-e^-$ spectrum) and $\mu^+-\mu^-$ (inducing a harder $e^+-e^-$ spectrum), and show exclusion plot in terms of $M_{\chi}$ vs $(\sigma_a v)$ (annihilating case) and $M_{\chi}$ vs $\tau$ (decaying case) in Figs.~\ref{fig:mvssv}b and \ref{fig:mvssv}c. 

Although the impact of various single assumptions is quite different in annihilating models with respect to decaying scenarios, there is roughly three orders of magnitude between the MIN and MAX constraints in both cases. 
Only for low mass WIMPs and fairly optimistic assumptions we can close on the benchmark `thermal' annihilation rate $(\sigma_a v)=3\cdot10^{-26}\, {\rm cm^3s^{-1}}$.
In the decaying case, the inclusion of angular power spectrum data leads to significant improvements with respect to constraints derived from total intensity, especially for large masses (while in the annihilating case the improvement is negligible). This can be understood also from Fig.~\ref{fig:cl}b.

\section{Conclusion}
\label{sec:conclusions}

In this paper we presented an extensive analysis on the cosmological radio emission associated to WIMP dark matter. This emission
is given by synchrotron radiation involving electrons and positrons injected by DM annihilations or decays in extragalactic structures.
WIMP models with large branching ratios into light leptons lead to large $e^+/e^-$ yields with suppressed $\pi^0$ production and in turn no bright $\gamma$--ray emission from $\pi^0$--decay. For those models the ratio between the DM signal and
the background astrophysical emission produced by ``standard'' sources is significantly higher at radio frequencies than in the $\gamma$--ray band.

For what concerns non--thermal emissions, radio telescopes have probably greater capabilities (i.e., sensitivity and angular resolution) as compared to telescopes at other wavelengths. This allows to study few different and complementary observables.
We considered the total intensity, the angular correlation of source distribution and anisotropies, and the source number counts.
We found that the emission induced by DM is mostly provided by a large number of small halos at redshift well below one. 
In general, WIMP models which provide a non--negligible contribution to the total extragalactic intensity take over other astrophysical sources in the number counts at low brightness ($S\lesssim \mu$Jy).
There are also interesting prospects for the angular correlation at low multipoles and, again, very low brightness. 
However, it is important to stress that such conclusions are valid under the assumption that the source population given by DM--induced signals can be treated as a separate population with respect to other astrophysical sources, as, e.g., AGN and SFG. On the other hand, any astrophysical source is embedded in a (typically much fainter) DM halo which makes the separation of the DM population awkward. 

We discussed the impact on the expected signals of dark matter clustering, magnetic fields, spatial diffusion, and dark matter microscopic properties. 
We found that bounds in the WIMP parameter space are uncertain by roughly three orders of magnitude.
Under optimistic assumptions, they close on light WIMPs with thermal annihilation cross section, and, in general, are competitive with constraints from other indirect dark matter searches.

The extragalactic background induced by DM can account for the ARCADE excess~\cite{Fornengo:2011cn} and the faint edge of differential source counts and angular correlation data, which will be provided by the next--generation radio surveys, can become an important probe for WIMPs.

\acknowledgments

NF and MR acknowledge research grants funded jointly by Ministero
dell'Istruzione, dell'Universit\`a e della Ricerca (MIUR), by
Universit\`a di Torino and by Istituto Nazionale di Fisica Nucleare
within the {\sl Astroparticle Physics Project} (MIUR contract number: PRIN 2008NR3EBK;
INFN grant code: FA51). RL, MT and NF acknowledge the spanish MICINN
Consolider Ingenio 2010 Programme under grant MULTIDARK CSD2009- 00064: NF for support,
RL and MT for support and for a fellowship. RL and MT were also supported by the EC contract UNILHC PITN-GA-2009-237920, by the Spanish grants FPA2008-00319 and PROMETEO/2009/091 (Generalitat Valenciana). The research of MT is partly supported by a Theory Fellowship of the Institute of Particle Physics (IPP)   and by the Natural Sciences and Engineering Research Council (NSERC) of Canada.

\end{document}